\documentclass[aps,prl,twocolumn,amsmath,longbibliography,superscriptaddress]{revtex4-2}  
\usepackage{bbm}
\usepackage{mathrsfs}
\usepackage{amsmath}
\usepackage{amsfonts}
\usepackage[colorlinks=true,citecolor=blue,anchorcolor=blue]{hyperref}
\usepackage{graphicx,epstopdf}
\usepackage{subfigure}
\usepackage{epsfig}
\usepackage{dcolumn}
\usepackage{bm}
\usepackage{color}
\usepackage{natbib}
\usepackage{amssymb}
\usepackage{xcolor}
\usepackage{braket}
\usepackage{array}
\usepackage{booktabs}
\usepackage{multirow}
\newcolumntype{C}[1]{>{\centering\arraybackslash}p{#1}}
\newcolumntype{L}[1]{>{\raggedright\arraybackslash}p{#1}}

\begin{document}

\author{Tongshuai Zhu}
\email{tongshuaizhu@upc.edu.cn}
\affiliation{College of Science, China University of Petroleum (East China), Qingdao 266580, China}
\affiliation{Eastern Institute of Technology, Ningbo 315200, China}

\author{Zixuan Li}
\affiliation{Eastern Institute of Technology, Ningbo 315200, China}

\author{Huaiqiang Wang}
\affiliation{Center for Quantum Transport and Thermal Energy Science, Institute of Physics Frontiers and Interdisciplinary Sciences, School of Physics and Technology, Nanjing Normal University, Nanjing 210023, China}
\affiliation{National Laboratory of Solid State Microstructures, Nanjing University, Nanjing 210093, China}
\affiliation{Jiangsu Physical Science Research Center, Nanjing 210093, China}
\author{Su-Huai Wei}
\affiliation{Eastern Institute of Technology, Ningbo 315200, China}

\author{Jiawei Ruan}
\email{jwruan@eitech.edu.cn}
\affiliation{Eastern Institute of Technology, Ningbo 315200, China}






\title{Floquet spin-group theory for light-tailored spin splitting in collinear magnets}

\begin{abstract}

Spin groups have emerged as a fundamental symmetry framework for spin-splitting antiferromagnets, including altermagnets. For periodically driven magnets, where new spin-splitting electronic structures can emerge beyond equilibrium, an analogous symmetry theory is still lacking. Here, we develop a Floquet spin-group theory for optically driven collinear magnets by combining the static spin group of the magnetic crystal with the dynamical symmetry group of the light field. This theory determines the symmetry-allowed parity, momentum dependence, and nodal structure of light-induced spin splitting under general periodic driving. We show that odd-, even-, and mixed-parity spin splittings can be selectively switched within the same material by tailoring the drive symmetry, accompanied by distinct symmetry-allowed transport responses. We further uncover higher-order $h$- and $k$-wave Floquet spin splittings unique to three-dimensional systems. Our work extends spin-group theory from static magnets to periodically driven systems, providing a general symmetry principle for engineering light-controlled nonequilibrium spin-resolved electronic structures.
\end{abstract}
\maketitle

\emph{Introduction}-Momentum-dependent spin splitting in compensated magnets has recently attracted considerable interest because it enables spin-polarized electronic responses without requiring a net magnetization, offering new opportunities for antiferromagnetic spintronics \cite{Libor2022Giant,Libor2022Beyond,Libor2022Emerging,krempasky2024altermagnetic,amin2024nanoscale,zhou2025manipulation,Lee2024Broken,Ding2024Large,Duan2025Antiferroelectric,Gu2025Ferroelectric,Wang2025Spin,Zhu2026Altermagnetic,song2025altermagnets,ma2021multifunctional,Karube2022Observation,regmi2025altermagnetism,han2024electrical,jiang2025metallic,zhang2025crystal,Liu2024Twisted,Lin2025Coulomb,Leeb2024Spontaneous,Wu2017Fermi,xt23-9pnv,fukaya2025superconducting,chen2026identifying}. A prominent example is altermagnetism, where collinear compensated magnets typically exhibit spin splitting with even parity under momentum inversion\cite{Libor2022Beyond}. More recently, odd-parity spin splitting in magnets has been identified beyond conventional altermagnets, emerging from both equilibrium unconventional orders \cite{song2025electrical,yamada2025metallic,Yu2025Odd,hellenes2023p,zhang2026quenching,Hayami2020Spontaneous,li2026p,zhuang2025odd,lin2025odd,leeb2026collinear} and nonequilibrium symmetry breaking induced by light \cite{Zhu2026Floquet,Huang2026Light,Liu2026Light,Li2026Floquet,tian2026optically,li2025robust}. Such odd-parity states can exhibit transport and superconducting-pairing properties distinct from those of even-parity magnets \cite{chakraborty2025highly,Brekke2024,Luo2026,khodas2026nonrelativistic}. Mixed-parity spin splitting, which combines even- and odd-parity contributions and therefore has no definite parity under momentum inversion, has also been proposed \cite{zhuang2026mixed}. These developments raise a natural question: can distinct spin-splitting symmetries be controllably selected and switched within the same material?

\begin{figure*}[t]
  \centering
  \includegraphics[width=\textwidth]{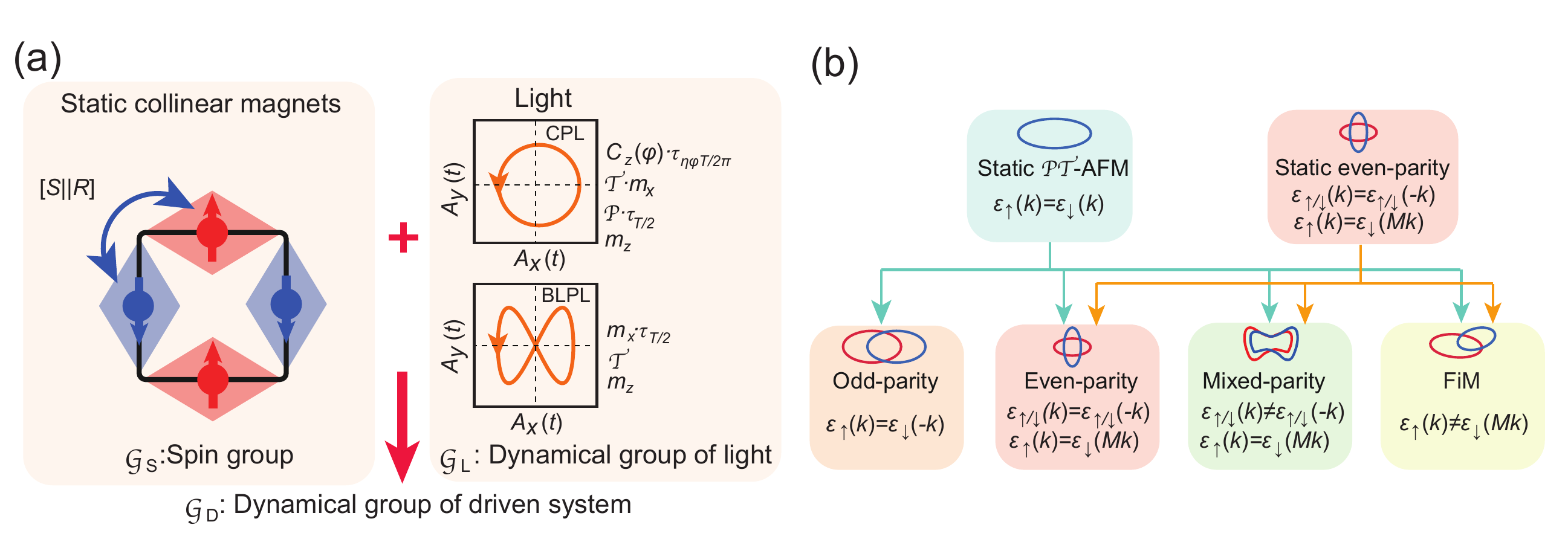}
  \caption{{Floquet spin splittings and dynamical symmetry in driven collinear magnets.} (a) In the negligible-SOC limit, the symmetries of  static collinear magnets are described by spin groups (left panel), whereas those of periodic driving fields are characterized by dynamical groups (right panel). For circularly polarized light (CPL) with vector potential $A_0(\cos \omega t,\sin \omega t,0)$ and bichromatic linearly polarized light (BLPL)  with vector potential $A_0(\sin \omega t,\sin 2\omega t,0)$, the generators of the corresponding dynamical groups are listed in the figure. The dynamical symmetry of the driven system is jointly determined by the spin group of the collinear magnets and the dynamical group of the driving field (See Eq.~(\ref{Eqcore}). (b) Schematic classification of light-induced spin-splitting patterns in collinear compensated magnets. Starting from a static $\mathcal{PT}$-symmetric antiferromagnet or an even-parity altermagnets, light driving can generate odd-parity, even-parity, mixed-parity, or ferrimagnetic (FiM)-like spin splitting, depending on the surviving dynamical symmetries. $M$ denotes a relevant spatial symmetry operation other than inversion $\mathcal{P}$.}\label{fig1}
\end{figure*}

In parallel, Floquet engineering has been widely used to reshape electronic structures and induce nonequilibrium phases \cite{wang2013observation,zhou2023pseudospin,mciver2020light,hubener2017creating,choi2025observation,bao2024manipulating,liu2025signatures,merboldt2025observation,oka2009Photovoltaic,Fu2026Floquet,yarmohammadi2026efficient,Fu2026Light,oka2009Photovoltaic}. Circularly polarized light (CPL), for example, breaks time-reversal symmetry and can generate anomalous Hall effect \cite{mciver2020light}, convert Dirac semimetals into Weyl phases \cite{hubener2017creating}, and induce odd-parity spin splitting in collinear antiferromagnets (AFM) \cite{Zhu2026Floquet,Huang2026Light,Liu2026Light,Li2026Floquet,tian2026optically,li2025robust}. Beyond such monochromatic driving, multichromatic fields, such as bichromatic linearly polarized light (BLPL), offer a substantially richer set of tunable dynamical symmetries involving spatial transformations, time reversal, and fractional-periodic time translations \cite{yu2021dynamical,Engelhardt2021Dynamical,neufeld2019floquet, Xu2018Space}.  Recent theoretical and experimental studies have shown that such structured driving fields can break the crystalline symmetries \cite{ikeda2023photocurrent,trevisan2022bicircular} as well as probe and engineer topological phases \cite{mitra2024light,tyulnev2024valleytronics,lesko2026optical,lesko2026probing}. These developments point to the possibility of controlling magnetic spin structures through the symmetry of light. What remains lacking, however, is a general framework that determines how the dynamical symmetry of an arbitrary periodic drive modifies the intrinsic spin symmetry of a magnetic crystal and constrains the resulting Floquet spin splitting.

In this work, we  develop a Floquet dynamical spin group description for periodically driven collinear magnets. By combining the static spin group of a magnetic crystal \cite{Xiao2024Spin,Jiang2024Enumeration,Chen2024Enumeration,jungwirth2026symmetry,zeng2026odd,luo2025spin} with the dynamical symmetry group of the light, we determine the spin symmetries that survive under periodic driving and derive the corresponding constraints on spin-resolved Floquet states. This framework provides a unified symmetry classification of light-induced spin splittings in both two and three dimensions. In particular, by tailoring the symmetry of the drive, odd-, even-, and mixed-parity spin splittings can be selectively switched within the same material class. We further show that CPL can induce higher-order  $h$-wave and $k$-wave spin splittings in three dimensions. Our work provides general principles for engineering nonequilibrium spin-split band structures in antiferromagnets and  altermagnets.



\emph{Dynamical spin point groups}-To simplify the analysis, here we focus on collinear magnetic order in the weak spin-orbit coupling (SOC) limit and restrict our discussion to the spin point group. The formulation can be  easily generalized to noncollinear magnets and to the full spin-space-group framework. An element of the static spin point group ${G}_{s}$ is denoted by $[S||R]\in {G}_{s}$, where $R$ is a spatial point-group operation and $S\in\{E,\mathcal{T},C_2,C_2\mathcal{T}\}\ltimes SO(2)$. Here,  $C_2$  represents a twofold spin rotation about an axis perpendicular to magnetic-moment direction, and $\mathcal{T}$ denotes time reversal, which is antiunitary in Hilbert space.  


In the absence of the driving field, the time-independent Bloch Hamiltonian is denoted by $H_{s}(\mathbf k)$, where $s=\pm1$ for spin-up/down. For a static spin point group operation $[S||R]$, the undriven Hamiltonian satisfies
\begin{equation}
\begin{split}
\hat{D}_{[S||R]}H_s(\mathbf k)\hat{D}_{[S||R]}^{-1}
=
H_{\chi_S s}(\eta_S R\mathbf k),
\end{split}
\label{EQ-static}
\end{equation}
where $\hat{D}_{[S||R]}$ is the representation of $[S||R]$,  and $\chi_S=\pm 1$ specifies whether the spin is preserved or reversed under the spin operation $S$.  $\eta_S=\pm 1$ distinguishes unitary spin operation (such as $C_2$ and $E$) from antiunitary ones (such as $C_2\mathcal{T}$ and $\mathcal{T}$).  For an antiunitary operator, the antiunitary character contributes an additional reversal of crystal momentum  $\textbf{k}$, and also reverses the time argument, $t\rightarrow-t$. The latter is immaterial for a static Hamiltonian and is therefore suppressed here, but must be retained for the time-dependent Hamiltonian considered below.

Before describing the symmetries of the light-driven time-periodic Hamiltonian, we first characterize the dynamical symmetry of the driving field through its vector potential $\mathbf A(t)$. In addition to the discrete time-translation symmetry $\mathbf{A}(t+T)=\mathbf{A}(t)$ with $T$ being the driving period, $\mathbf{A}(t)$ may also be invariant under combined spatial and temporal transformations Ref. \cite{neufeld2019floquet}.  We consider an operation  $ g_L =  \Theta  \cdot  R  \cdot \tau_\delta $, where $R$ is a point-group operation, $\tau_\delta:t\mapsto t-\delta$  is a time translation, and $\Theta \in \{\mathcal{E},\mathcal{T}\} $ with  $\mathcal{E}$ the temporal identity operator and $\mathcal{T}$ the time reversal. We will suppress $\mathcal{E}$ when no ambiguity arises. Applying $g_L$  on the vector potential leads to $ g_L \mathbf{A}(t)=\eta_\Theta R\mathbf{A}(\eta_\Theta t-\eta_\Theta\delta)$. Here, $\eta_\Theta=1$ for $\Theta=\mathcal{E}$ while $\eta_\Theta=-1$ for $\Theta=\mathcal{T}$, reflecting the odd transformation of the vector potential and the time argument under time reversal. The dynamical  symmetry group of the light field is therefore  
\begin{equation}
\mathcal{G}_{\mathrm{L}}=\left\{\Theta \cdot R\cdot \tau_\delta \mid \mathbf{A}(t)=\eta_{\Theta} R \mathbf{A}\left(\eta_{\Theta} (t-\delta )\right)\right\}.
\label{eq:light_symm}
\end{equation}

For the driven Hamiltonian $H_s(\mathbf k,t)=H_s[\mathbf k+e\mathbf A(t)/\hbar]$, the vector potential $\mathbf{A}(t)$ couples to the crystal momenta through the Peierls substitution $\mathbf k\rightarrow\mathbf k+e\mathbf A(t)/\hbar$ and therefore acts directly on the orbital rather than the spin sector. Hence, a spin-group operation $[S||R]$ can  remain a symmetry under driving only if its corresponding orbital transformation is compatible with a dynamical symmetry of the light field.  This motivates the definition of the \textit{dynamical spin point group} for the light-driven magnets as 
\begin{equation}
\mathcal{G}_\mathrm{D}=
\left\{
[S\Vert R]\cdot\tau_\delta
\;\middle|\;
\begin{aligned}
&[S\Vert R]\in G_s,\\[2pt]
&\pi([S\Vert R])\cdot \tau_\delta\in \mathcal{G}_\mathrm{L}
\end{aligned}
\right\}.
\label{Eqcore}
\end{equation}
Here,  $\pi$ denotes the map from a spin-group operation to its corresponding orbital transformation, with $\pi([S\Vert R])=R$ for unitary spin operator $S$ and $\pi([S\Vert R])=\mathcal{T}\cdot R$ for antiunitary spin operator $S$. Eq. (\ref{Eqcore}) thus selects each $[S\Vert R] \in G_s$ together with a time translation $\tau_\delta$ for which $\pi([S\Vert R])\cdot \tau_\delta$ is a symmetry of the driving field. For each element $\gamma=[S\Vert R]\cdot\tau_\delta \in \mathcal{G}_{\mathrm{D}}$, whose action is represented as $\hat{D}_\gamma = \hat{D}_{[S||R]} \cdot \tau_\delta $ , the time-dependent Hamiltonian satisfies
\begin{equation}
\hat{D}_\gamma H_s\left(\mathbf{k}, t\right) \hat{D}_\gamma^{-1} =H_{\chi_S s}\left(\eta_S R \mathbf{k}, t\right)
\label{Eq-invariant}
\end{equation}
which has the similar form as the corresponding symmetry relation for the static Hamiltonian in Eq.~(\ref{EQ-static}) [see Eq.~(\ref{transformation_Dg}) in the End Matter for the derivation].

\begin{figure*}[t]
  \centering
  \includegraphics[width=0.8\textwidth]{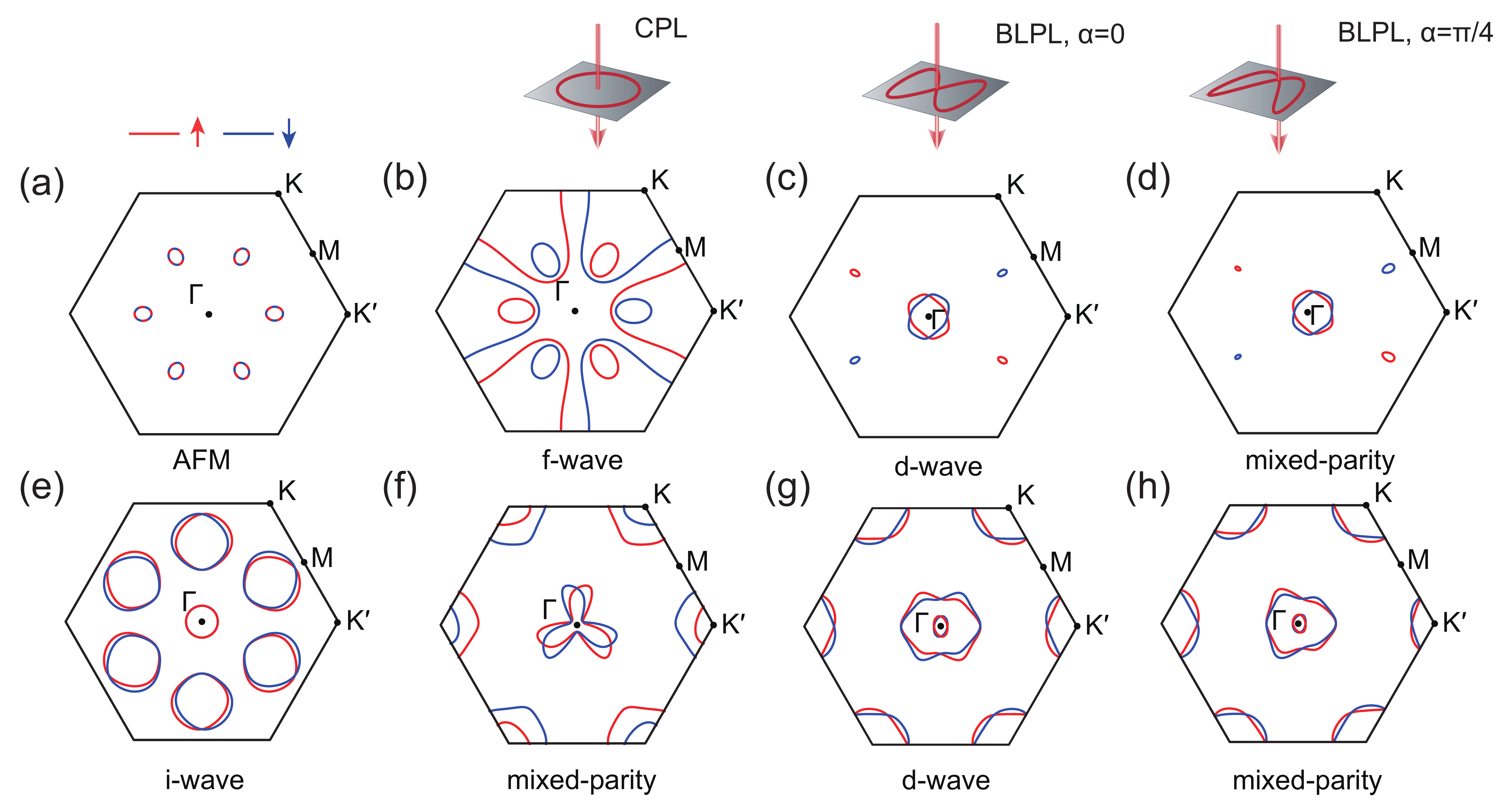}
  \caption{{ Light-selected spin splitting in 2D collinear magnets.} ({a-d}) Calculated constant-energy contours of MnPSe$_3$ (a) in the absence of light, (b) under CPL irradiation, (c) under BLPL irradiation with $\alpha=0$ and (d) under BLPL irradiation with $\alpha=\pi/4$. The corresponding  constant energies are $E_F=-0.3$ eV for (a,b) and $E_F=-0.35$ eV for (c,d). The driving parameters are $\hbar\omega=10$ eV and $\tilde A=0.35$ \AA$^{-1}$. (e-h) Same as (a-d) for Mn$_2$P$_2$S$_3$Se$_3$ , with $E_F=-0.2$ eV in (e) and $E_F=-0.335$ eV in (f-h).  The driving parameters for (f-h) are $\hbar\omega=10$ eV and $\tilde A=0.25$ \AA$^{-1}$.}\label{fig2}
\end{figure*}

We now examine how these symmetries constrain the quasienergy spectrum. The Floquet quasienergies are obtained from
\begin{equation}
 H^F_s(\mathbf k,t) |u_{s,\beta}(\mathbf k,t)\rangle=\varepsilon_{s,\beta}(\mathbf k) |u_{s,\beta}(\mathbf k,t)\rangle,
 \label{EQ-floquetEQ}
 \end{equation}
where $H^F_s(\mathbf k,t)=H_s(\mathbf k,t)-i\hbar\partial_t$ is the Floquet Hamiltonian, $|u_{s,\beta}(\mathbf k,t)\rangle$ is the Floquet mode labeled by $\beta$, satisfying $|u_{s,\beta}(\mathbf k,t)\rangle=|u_{s,\beta}(\mathbf k,t+T)\rangle$, and $\varepsilon_{s,\beta}(\mathbf k)$ is the  corresponding quasienergy. By utilizing the relation in Eq.~(\ref{Eq-invariant}) and after some algebra (see the End Matter), we find that for a  dynamical spin-group operation $\gamma=[S\Vert R]\cdot\tau_\delta \in \mathcal{G}_{\mathrm{D}}$, for the non-degenerate band, the quasienergy $\varepsilon_{s,\beta}(\mathbf k)$ obeys
\begin{equation}
\varepsilon_{\chi_S s,\beta}(\eta_S R\mathbf k)
=
\varepsilon_{s,\beta}(\mathbf k)
\quad (\mathrm{mod}\ \hbar\omega).
\label{EQ-energy}
\end{equation}
Remarkably, the constraint on the quasienergy spectrum depends only on the spin-group part. Time translation $\tau_\delta$ , however, remains encoded in the transformation of the Floquet modes and therefore affects their time-dependent wavefunctions  (See Eq.~(\ref{wavefunction_time}) in the End Matter). 

Equation~(\ref{EQ-energy}) provides a simple route to analyzing light-induced spin-splitting patterns by examining the surviving spin-group symmetries defined in Eq.~(\ref{Eqcore}). Based on this principle, we classify possible spin-splitting patterns into four types: (i) even-parity spin splitting (i.e. AM spin splitting), where opposite-spin sectors are related at symmetry-related momenta, $\varepsilon_{\uparrow,\beta}(\eta_S R\mathbf k)=\varepsilon_{\downarrow,\beta}(\mathbf k)$, while each spin sector satisfies $\varepsilon_{s,\beta}(\mathbf k)=\varepsilon_{s,\beta}(-\mathbf k)$; (ii) odd-parity splitting, characterized by, $\varepsilon_{\uparrow,\beta}(\mathbf k)=\varepsilon_{\downarrow,\beta}(-\mathbf k)$; (iii) mixed-parity spin splitting, for which neither even- nor odd-parity symmetry holds, although a spin-flip symmetry still relates the two spin sectors, $\varepsilon_{\uparrow,\beta}(\eta_S R\mathbf k)=\varepsilon_{\downarrow,\beta}(\mathbf k)$; and (iv), ferrimagnetic-like spin-splitting, where no spin-point-group operation relates the two spin sectors.  We note that cases (ii) and (iii) are usually absent in equilibrium, and by tailoring the symmetry of the light, one can generate and adjust the four cases on demand. 

\emph{Classification of light-induced spin splittings in 2D}-In the following, we consider CPL and BLPL drivings with light propagating normal to the 2D plane. For CPL, the vector potential  $\mathbf A^{\text{CPL}}(t)=A_0(\cos{\omega t},\eta\sin{\omega t},0)$, whose dynamical symmetry group is generated by $C_z(\varphi)\cdot \tau_{\eta\varphi T/2\pi}$,  $\mathcal{T}\cdot m_{x}$, $m_z$ and $\mathcal{P}\cdot\tau_{T/2}$, where $C_z(\varphi)$ denotes a rotation about the $z$ axis by an angle $\varphi$,  $\eta=\pm1$ denotes the helicity, $m_{x}$ is a vertical mirror reflection, and $\mathcal{P}$ is the inversion.  We also consider BLPL with $\mathbf A^{\text{BLPL}}_{\alpha}(t)=A_0(\sin\omega t,\sin(2\omega t+\alpha),0)$, where $\alpha$ is the relative phase between the fundamental and second-harmonic components.  For $\alpha=0,\pi$, the dynamical group is generated by $m_{x}\cdot \tau_{T/2}$, $m_z$ and  $\mathcal{T}$, whereas for a generic  $\alpha\neq\pi,0,\pm\pi/2$,  the symmetry $m_{x}\cdot\tau_{T/2}$ and $m_z$ remains (See Sec.~S2 of the SM for more details). 




We have systematically examined all 63 spin-layer point groups of 2D compensated collinear magnets \cite{Wang2026Type} under the three driving protocols considered above. The surviving spin group symmetries of each case are listed in the SM, and the resulting spin-splitting types are summarized in Table~\ref{tab:2d}. Notably, for an undriven AFM possessing the symmetry $[C_2||m_{001}]$, the mirror operation $m_{001}$ leaves the in-plane momentum $\mathbf{k}$ invariant while $C_2$ flips the spin. This symmetry is preserved by all normally incident driving fields and therefore continues to enforce spin degeneracy, as reflected in the first row of Table~\ref{tab:2d}.  


Intriguingly, in some systems, the symmetry of the driving field can selectively switch the spin splitting among odd-, even-, and mixed-parity forms within the same material. As a demonstration, we performed first-principles calculations of the constant-energy contours of two representative materials MnPSe$_3$ and  Mn$_2$P$_2$S$_3$Se$_3$, under different light fields. The spin point group for monolayer MnPSe$_3$ is $\overline{ }^{\overline{1}} \overline{3}^{\overline{1}} m^{\infty m} 1$ (No.~46).  According to Table~\ref{tab:2d},  in the absence of light, the constant-energy contours is spin degenerate [Fig.~\ref{fig2}(a)].  CPL converts the spin degenerate bands into an $f$-wave spin splitting [Fig.~\ref{fig2}(b)]. By contrast, BLPL with $\alpha=0$  preserves  $[C_2\mathcal{T}||E]$, $[C_2||m_{120}]$ and while breaking $[C_2||\mathcal{P}]$, thereby producing an even-parity $d$-wave spin splitting  [Fig.~\ref{fig2}(c)]. Tuning the relative phase to $\alpha=\pi/4$ further breaks   $[C_2\mathcal{T}||E]$, leaving only $[C_2||m_{120}]$ as the spin flip symmetry and giving rise to a mixed-parity spin splitting [Fig.~\ref{fig2}(d)].  We note that the above spin-splitting parities are associated with distinct symmetry constraints and can therefore give rise to different transport responses, such as anomalous Hall effects and Berry-curvature-dipole induced nonlinear hall effects (See Sec. S3 of the SM).

Mn$_2$P$_2$S$_3$Se$_3$ provides a representative example of a system that is already altermagnetic in equilibrium\cite{mazin2023induced}. Its spin point group is ${ }^1 3^{\overline{1}} m^{\infty m} 1$ and the undriven state exhibits an $i$-wave spin texture [Fig.~\ref{fig2}(e)]. Under CPL irradiation, $[C_2\mathcal{T}||E]$ is broken, while  $[E\|3_{001}]$ and $[\mathcal{T}\|m_{120}]$ remain preserved. As a result, $\mathbf{k}$ and $-\mathbf{k}$ are no longer related by any symmetry, and the constant-energy contours develops a mixed-parity spin splitting [Fig.~\ref{fig2}(f)]. By contrast, BLPL with $\alpha=0$  selects a $d$-wave spin-splitting pattern [Fig.~\ref{fig2}(g)]. When the relative phase is tuned to $\alpha=\pi/4$ only  $[C_2||m_{120}]$ remains as a spin-flip symmetry,  leading to a mixed-parity spin splitting [Fig.~\ref{fig2}(h)].

\emph{Light-induced spin splittings in 3D}- Several symmetries that enforce spin degeneracy in 2D no longer do so in 3D. A representative example is the spin group operation $[C_2||m_{001}]$. In a 2D system, the mirror operation $m_{001}$ acts trivially on the in-plane crystal momentum, leaving each momentum point invariant. Consequently, $[C_2||m_{001}]$ relates two opposite spin states at the same $\mathbf{k}$, thereby enforcing spin degeneracy throughout the Brillouin zone. For a 3D momentum $\mathbf{k}=(\mathbf{k}_{\parallel},k_z)$, the same mirror operation maps $(\mathbf{k}_{\parallel},k_z)$ to $(\mathbf{k}_{\parallel},-k_z)$.  This dimensional distinction allows CPL to induce spin splitting in most 3D spin-degenerate AFM systems. The corresponding spin-splitting types are summarized in Table~\ref{tab:3d} in the End Matter. We further find that 3D systems can host higher-order odd-parity spin splittings that are absent in 2D, as will be demonstrated below.

\begin{figure*}[t]
  \centering
  \includegraphics[width=0.8\textwidth]{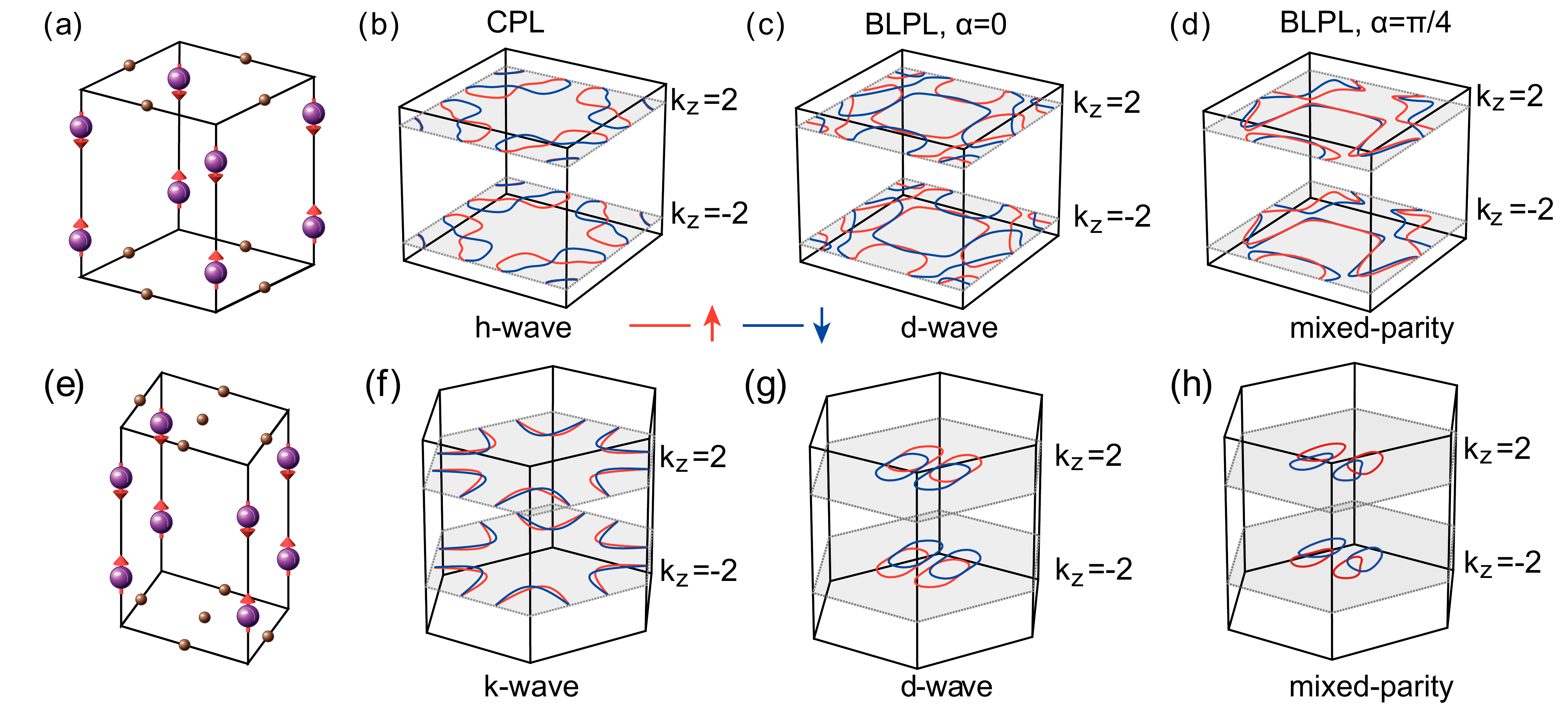}
  \caption{Light-selected spin splitting in 3D collinear magnets. (a) Tetragonal lattice model with spin point group${ }^1 4 /^{\overline{1}} m^1 m^1 m^{\infty m} 1$. (b-d) Constant-energy contours at   $E_F=1.2$ eV  under CPL irradiation (b), under BLPL irradiation with $\alpha=0$ (c), and under  BLPL irradiation with $\alpha=\pi/4$ (d). (e) Hexagonal lattice model  with spin point group ${}^1 6/{}^{\bar1}m{}^1m{}^1m{}^{\infty m}1$. (f-h) Constant-energy contours of the hexagonal model at $E_F=-2.4$ eV under CPL irradiation (f), under BLPL irradiation with $\alpha=0$ (g) and under  BLPL irradiation with $\alpha=\pi/4$ (h). For both models, the driving parameters are $\hbar\omega=10$ eV and $\tilde A a=2$.}\label{fig3}
\end{figure*}

As a demonstration, we construct representative 3D tight-binding models (See the SM for more details \cite{SM}). As a first example, we consider a lattice  respecting the spin point group symmetry ${ }^1 4 /^{\overline{1}} m^1 m^1 m^{\infty m} 1$, as shown in Fig.~\ref{fig3}(a). The magnetic atoms are placed at the Wyckoff position $2e$, while the nonmagnetic atoms are placed at the Wyckoff position $2f$. We assign one  orbital to each site and construct the corresponding 3D tight-binding Hamiltonian. As shown in Fig.~\ref{fig3}(b),  fixed-$k_z$ cuts of the  constant-energy contours, calculated within the high-frequency approximation under CPL irradiation, exhibit the expected in-plane $g$-wave pattern. Moreover,  the spin polarization changes sign between the mirror-related planes $k_z$ and -$k_z$, confirming the full 3D $h$-wave spin splitting (see Sec.~S6 of the SM for more details). Under BLPL with $\alpha=0$, the breaking of $[C_2||\mathcal{P}]$ produces a $d$-wave spin splitting [Fig.~\ref{fig3}(c)], whereas BLPL with $\alpha=\pi/4$ breaks both $[C_2||\mathcal{P}]$ and $[C_2\mathcal{T}||E]$, giving rise to a mixed-parity spin texture [Fig.~\ref{fig3}(d)].

The second model belongs to the spin point group ${}^{1}6/{}^{\overline1}m{}^{1}m{}^{1}m{}^{\infty m}1$, with magnetic atoms occupying the Wyckoff position $2e$ and nonmagnetic atoms the $3f$ position [Fig.~\ref{fig3}(e)]. Fixed-$k_z$ cuts of the constant-energy contours exhibit the characteristic in-plane $i$-wave pattern, with the spin polarization reversing between the mirror-related planes $k_z$ and $-k_z$ [Fig.~\ref{fig3}(f)], giving rise to the full 3D $k$-wave texture.  BLPL with $\alpha=0$ produces a $d$-wave spin splitting [Fig.~\ref{fig3}(g)], whereas $\alpha=\pi/4$ breaks both $[C_2||\mathcal{P}]$ and $[C_2\mathcal{T}||E]$, leading to a mixed-parity spin texture [Fig.~\ref{fig3}(h)].

\emph{Discussion}-The predicted light-induced spin splittings can be directly probed by spin- and time-resolved ARPES. In Sec.~S5 of the SM, we first calculate the Floquet-only photoemission spectra, which clearly resolve the predicted spin splitting. We then incorporate Floquet–Volkov interference between the light-dressed initial states and Volkov final states to model the photoemission process more realistically.  Remarkably, the resulting spectra not only present the signatures of the spin splitting but also retain the fractional time translations associated with the dynamical symmetry. Although these time translations do not alter the Floquet quasienergies, they imprint harmonic-dependent phases on the Floquet states [Eq.~(\ref{wavefunction_time}) in the End Matter]. Through Floquet–Volkov interference, these phase factors are converted into spectral-weight modulations, leading to asymmetric photoemission intensities between symmetry-related Floquet states (see Sec.~S5 of the SM for more details).


\emph{Conclusion}-We have developed a Floquet spin-group framework that combines the static spin symmetry of a collinear magnet with the dynamical symmetry of a periodic driving field. The resulting symmetry relations provide direct constraints on spin-resolved Floquet quasienergies and a unified classification of light-induced spin splittings. By tailoring the dynamical symmetry of the drive, distinct spin-splitting patterns can be selectively switched within the same material, while three-dimensional systems further allow higher-order $h$- and $k$-wave textures. Although our work focus on periodically driven collinear magnets, the symmetry framework is general and can be extended to study driven noncollinear and noncoplanar magnets. Our study provides a symmetry-based route to controlling spin-resolved Floquet band structures and their associated transport responses with light.

\emph{Note added}-We recently became aware of two independent and concurrent studies discussing spin splittings with higher-order symmetry in three-dimensional systems\cite{liu2026complete}, as well as the switching between odd- and even-parity spin splittings in coplanar magnets\cite{zhu2026parity}.

\emph{Acknowledgments}-This work is supported by the National Natural Science Foundation of China (Grants No. 12534007, No. 12504197, No. 12504280, No. 12688201), the National Key Research and Development Program of China (Grant No. 2024YFA1409800), Shandong Provincial Natural Science Foundation (Grant No. ZR2024QA095), the  Fundamental Research Funds for the Central Universities  (Grant No. 23CX06063A), the Natural Science Foundation  of Jiangsu Province (Grants No. BK20252117,  No. BK20233001), the Youth Innovation Team Plan  Project for the Higher Education Institution of Shandong  Province (Grant No. 2024KJN021).

\bibliographystyle{apsrev4-2}
\bibliography{ref}

\begin{thebibliography}{90}%
\makeatletter
\providecommand \@ifxundefined [1]{%
 \@ifx{#1\undefined}
}%
\providecommand \@ifnum [1]{%
 \ifnum #1\expandafter \@firstoftwo
 \else \expandafter \@secondoftwo
 \fi
}%
\providecommand \@ifx [1]{%
 \ifx #1\expandafter \@firstoftwo
 \else \expandafter \@secondoftwo
 \fi
}%
\providecommand \natexlab [1]{#1}%
\providecommand \enquote  [1]{``#1''}%
\providecommand \bibnamefont  [1]{#1}%
\providecommand \bibfnamefont [1]{#1}%
\providecommand \citenamefont [1]{#1}%
\providecommand \href@noop [0]{\@secondoftwo}%
\providecommand \href [0]{\begingroup \@sanitize@url \@href}%
\providecommand \@href[1]{\@@startlink{#1}\@@href}%
\providecommand \@@href[1]{\endgroup#1\@@endlink}%
\providecommand \@sanitize@url [0]{\catcode `\\12\catcode `\$12\catcode
  `\&12\catcode `\#12\catcode `\^12\catcode `\_12\catcode `\%12\relax}%
\providecommand \@@startlink[1]{}%
\providecommand \@@endlink[0]{}%
\providecommand \url  [0]{\begingroup\@sanitize@url \@url }%
\providecommand \@url [1]{\endgroup\@href {#1}{\urlprefix }}%
\providecommand \urlprefix  [0]{URL }%
\providecommand \Eprint [0]{\href }%
\providecommand \doibase [0]{https://doi.org/}%
\providecommand \selectlanguage [0]{\@gobble}%
\providecommand \bibinfo  [0]{\@secondoftwo}%
\providecommand \bibfield  [0]{\@secondoftwo}%
\providecommand \translation [1]{[#1]}%
\providecommand \BibitemOpen [0]{}%
\providecommand \bibitemStop [0]{}%
\providecommand \bibitemNoStop [0]{.\EOS\space}%
\providecommand \EOS [0]{\spacefactor3000\relax}%
\providecommand \BibitemShut  [1]{\csname bibitem#1\endcsname}%
\let\auto@bib@innerbib\@empty
\bibitem [{\citenamefont {\ifmmode~\check{S}\else \v{S}\fi{}mejkal}\ \emph
  {et~al.}(2022{\natexlab{a}})\citenamefont {\ifmmode~\check{S}\else
  \v{S}\fi{}mejkal}, \citenamefont {Hellenes}, \citenamefont
  {Gonz\'alez-Hern\'andez}, \citenamefont {Sinova},\ and\ \citenamefont
  {Jungwirth}}]{Libor2022Giant}%
  \BibitemOpen
  \bibfield  {author} {\bibinfo {author} {\bibfnamefont {L.}~\bibnamefont
  {\ifmmode~\check{S}\else \v{S}\fi{}mejkal}}, \bibinfo {author} {\bibfnamefont
  {A.~B.}\ \bibnamefont {Hellenes}}, \bibinfo {author} {\bibfnamefont
  {R.}~\bibnamefont {Gonz\'alez-Hern\'andez}}, \bibinfo {author} {\bibfnamefont
  {J.}~\bibnamefont {Sinova}},\ and\ \bibinfo {author} {\bibfnamefont
  {T.}~\bibnamefont {Jungwirth}},\ }\href
  {https://doi.org/10.1103/PhysRevX.12.011028} {\bibfield  {journal} {\bibinfo
  {journal} {Phys. Rev. X}\ }\textbf {\bibinfo {volume} {12}},\ \bibinfo
  {pages} {011028} (\bibinfo {year} {2022}{\natexlab{a}})}\BibitemShut
  {NoStop}%
\bibitem [{\citenamefont {\ifmmode~\check{S}\else \v{S}\fi{}mejkal}\ \emph
  {et~al.}(2022{\natexlab{b}})\citenamefont {\ifmmode~\check{S}\else
  \v{S}\fi{}mejkal}, \citenamefont {Sinova},\ and\ \citenamefont
  {Jungwirth}}]{Libor2022Beyond}%
  \BibitemOpen
  \bibfield  {author} {\bibinfo {author} {\bibfnamefont {L.}~\bibnamefont
  {\ifmmode~\check{S}\else \v{S}\fi{}mejkal}}, \bibinfo {author} {\bibfnamefont
  {J.}~\bibnamefont {Sinova}},\ and\ \bibinfo {author} {\bibfnamefont
  {T.}~\bibnamefont {Jungwirth}},\ }\href
  {https://doi.org/10.1103/PhysRevX.12.031042} {\bibfield  {journal} {\bibinfo
  {journal} {Phys. Rev. X}\ }\textbf {\bibinfo {volume} {12}},\ \bibinfo
  {pages} {031042} (\bibinfo {year} {2022}{\natexlab{b}})}\BibitemShut
  {NoStop}%
\bibitem [{\citenamefont {\ifmmode~\check{S}\else \v{S}\fi{}mejkal}\ \emph
  {et~al.}(2022{\natexlab{c}})\citenamefont {\ifmmode~\check{S}\else
  \v{S}\fi{}mejkal}, \citenamefont {Sinova},\ and\ \citenamefont
  {Jungwirth}}]{Libor2022Emerging}%
  \BibitemOpen
  \bibfield  {author} {\bibinfo {author} {\bibfnamefont {L.}~\bibnamefont
  {\ifmmode~\check{S}\else \v{S}\fi{}mejkal}}, \bibinfo {author} {\bibfnamefont
  {J.}~\bibnamefont {Sinova}},\ and\ \bibinfo {author} {\bibfnamefont
  {T.}~\bibnamefont {Jungwirth}},\ }\href
  {https://doi.org/10.1103/PhysRevX.12.040501} {\bibfield  {journal} {\bibinfo
  {journal} {Phys. Rev. X}\ }\textbf {\bibinfo {volume} {12}},\ \bibinfo
  {pages} {040501} (\bibinfo {year} {2022}{\natexlab{c}})}\BibitemShut
  {NoStop}%
\bibitem [{\citenamefont {Krempask{\`y}}\ \emph {et~al.}(2024)\citenamefont
  {Krempask{\`y}}, \citenamefont {{\v{S}}mejkal}, \citenamefont {D’souza},
  \citenamefont {Hajlaoui}, \citenamefont {Springholz}, \citenamefont
  {Uhl{\'\i}{\v{r}}ov{\'a}}, \citenamefont {Alarab}, \citenamefont
  {Constantinou}, \citenamefont {Strocov}, \citenamefont {Usanov} \emph
  {et~al.}}]{krempasky2024altermagnetic}%
  \BibitemOpen
  \bibfield  {author} {\bibinfo {author} {\bibfnamefont {J.}~\bibnamefont
  {Krempask{\`y}}}, \bibinfo {author} {\bibfnamefont {L.}~\bibnamefont
  {{\v{S}}mejkal}}, \bibinfo {author} {\bibfnamefont {S.}~\bibnamefont
  {D’souza}}, \bibinfo {author} {\bibfnamefont {M.}~\bibnamefont {Hajlaoui}},
  \bibinfo {author} {\bibfnamefont {G.}~\bibnamefont {Springholz}}, \bibinfo
  {author} {\bibfnamefont {K.}~\bibnamefont {Uhl{\'\i}{\v{r}}ov{\'a}}},
  \bibinfo {author} {\bibfnamefont {F.}~\bibnamefont {Alarab}}, \bibinfo
  {author} {\bibfnamefont {P.}~\bibnamefont {Constantinou}}, \bibinfo {author}
  {\bibfnamefont {V.}~\bibnamefont {Strocov}}, \bibinfo {author} {\bibfnamefont
  {D.}~\bibnamefont {Usanov}}, \emph {et~al.},\ }\href
  {https://doi.org/10.1038/s41586-023-06907-7} {\bibfield  {journal} {\bibinfo
  {journal} {Nature}\ }\textbf {\bibinfo {volume} {626}},\ \bibinfo {pages}
  {517} (\bibinfo {year} {2024})}\BibitemShut {NoStop}%
\bibitem [{\citenamefont {Amin}\ \emph {et~al.}(2024)\citenamefont {Amin},
  \citenamefont {Dal~Din}, \citenamefont {Golias}, \citenamefont {Niu},
  \citenamefont {Zakharov}, \citenamefont {Fromage}, \citenamefont {Fields},
  \citenamefont {Heywood}, \citenamefont {Cousins}, \citenamefont {Maccherozzi}
  \emph {et~al.}}]{amin2024nanoscale}%
  \BibitemOpen
  \bibfield  {author} {\bibinfo {author} {\bibfnamefont {O.}~\bibnamefont
  {Amin}}, \bibinfo {author} {\bibfnamefont {A.}~\bibnamefont {Dal~Din}},
  \bibinfo {author} {\bibfnamefont {E.}~\bibnamefont {Golias}}, \bibinfo
  {author} {\bibfnamefont {Y.}~\bibnamefont {Niu}}, \bibinfo {author}
  {\bibfnamefont {A.}~\bibnamefont {Zakharov}}, \bibinfo {author}
  {\bibfnamefont {S.}~\bibnamefont {Fromage}}, \bibinfo {author} {\bibfnamefont
  {C.}~\bibnamefont {Fields}}, \bibinfo {author} {\bibfnamefont
  {S.}~\bibnamefont {Heywood}}, \bibinfo {author} {\bibfnamefont
  {R.}~\bibnamefont {Cousins}}, \bibinfo {author} {\bibfnamefont
  {F.}~\bibnamefont {Maccherozzi}}, \emph {et~al.},\ }\href
  {https://doi.org/10.1038/s41586-024-08234-x} {\bibfield  {journal} {\bibinfo
  {journal} {Nature}\ }\textbf {\bibinfo {volume} {636}},\ \bibinfo {pages}
  {348} (\bibinfo {year} {2024})}\BibitemShut {NoStop}%
\bibitem [{\citenamefont {Zhou}\ \emph {et~al.}(2025)\citenamefont {Zhou},
  \citenamefont {Cheng}, \citenamefont {Hu}, \citenamefont {Chu}, \citenamefont
  {Bai}, \citenamefont {Han}, \citenamefont {Liu}, \citenamefont {Pan},\ and\
  \citenamefont {Song}}]{zhou2025manipulation}%
  \BibitemOpen
  \bibfield  {author} {\bibinfo {author} {\bibfnamefont {Z.}~\bibnamefont
  {Zhou}}, \bibinfo {author} {\bibfnamefont {X.}~\bibnamefont {Cheng}},
  \bibinfo {author} {\bibfnamefont {M.}~\bibnamefont {Hu}}, \bibinfo {author}
  {\bibfnamefont {R.}~\bibnamefont {Chu}}, \bibinfo {author} {\bibfnamefont
  {H.}~\bibnamefont {Bai}}, \bibinfo {author} {\bibfnamefont {L.}~\bibnamefont
  {Han}}, \bibinfo {author} {\bibfnamefont {J.}~\bibnamefont {Liu}}, \bibinfo
  {author} {\bibfnamefont {F.}~\bibnamefont {Pan}},\ and\ \bibinfo {author}
  {\bibfnamefont {C.}~\bibnamefont {Song}},\ }\href
  {https://doi.org/10.1038/s41586-024-08436-3} {\bibfield  {journal} {\bibinfo
  {journal} {Nature}\ }\textbf {\bibinfo {volume} {638}},\ \bibinfo {pages}
  {645} (\bibinfo {year} {2025})}\BibitemShut {NoStop}%
\bibitem [{\citenamefont {Lee}\ \emph {et~al.}(2024)\citenamefont {Lee},
  \citenamefont {Lee}, \citenamefont {Jung}, \citenamefont {Jung},
  \citenamefont {Kim}, \citenamefont {Lee}, \citenamefont {Seok}, \citenamefont
  {Kim}, \citenamefont {Park}, \citenamefont {\ifmmode~\check{S}\else
  \v{S}\fi{}mejkal}, \citenamefont {Kang},\ and\ \citenamefont
  {Kim}}]{Lee2024Broken}%
  \BibitemOpen
  \bibfield  {author} {\bibinfo {author} {\bibfnamefont {S.}~\bibnamefont
  {Lee}}, \bibinfo {author} {\bibfnamefont {S.}~\bibnamefont {Lee}}, \bibinfo
  {author} {\bibfnamefont {S.}~\bibnamefont {Jung}}, \bibinfo {author}
  {\bibfnamefont {J.}~\bibnamefont {Jung}}, \bibinfo {author} {\bibfnamefont
  {D.}~\bibnamefont {Kim}}, \bibinfo {author} {\bibfnamefont {Y.}~\bibnamefont
  {Lee}}, \bibinfo {author} {\bibfnamefont {B.}~\bibnamefont {Seok}}, \bibinfo
  {author} {\bibfnamefont {J.}~\bibnamefont {Kim}}, \bibinfo {author}
  {\bibfnamefont {B.~G.}\ \bibnamefont {Park}}, \bibinfo {author}
  {\bibfnamefont {L.}~\bibnamefont {\ifmmode~\check{S}\else \v{S}\fi{}mejkal}},
  \bibinfo {author} {\bibfnamefont {C.-J.}\ \bibnamefont {Kang}},\ and\
  \bibinfo {author} {\bibfnamefont {C.}~\bibnamefont {Kim}},\ }\href
  {https://doi.org/10.1103/PhysRevLett.132.036702} {\bibfield  {journal}
  {\bibinfo  {journal} {Phys. Rev. Lett.}\ }\textbf {\bibinfo {volume} {132}},\
  \bibinfo {pages} {036702} (\bibinfo {year} {2024})}\BibitemShut {NoStop}%
\bibitem [{\citenamefont {Ding}\ \emph {et~al.}(2024)\citenamefont {Ding},
  \citenamefont {Jiang}, \citenamefont {Chen}, \citenamefont {Tao},
  \citenamefont {Liu}, \citenamefont {Li}, \citenamefont {Liu}, \citenamefont
  {Sun}, \citenamefont {Cheng}, \citenamefont {Liu}, \citenamefont {Yang},
  \citenamefont {Zhang}, \citenamefont {Deng}, \citenamefont {Jing},
  \citenamefont {Huang}, \citenamefont {Shi}, \citenamefont {Ye}, \citenamefont
  {Qiao}, \citenamefont {Wang}, \citenamefont {Guo}, \citenamefont {Feng},\
  and\ \citenamefont {Shen}}]{Ding2024Large}%
  \BibitemOpen
  \bibfield  {author} {\bibinfo {author} {\bibfnamefont {J.}~\bibnamefont
  {Ding}}, \bibinfo {author} {\bibfnamefont {Z.}~\bibnamefont {Jiang}},
  \bibinfo {author} {\bibfnamefont {X.}~\bibnamefont {Chen}}, \bibinfo {author}
  {\bibfnamefont {Z.}~\bibnamefont {Tao}}, \bibinfo {author} {\bibfnamefont
  {Z.}~\bibnamefont {Liu}}, \bibinfo {author} {\bibfnamefont {T.}~\bibnamefont
  {Li}}, \bibinfo {author} {\bibfnamefont {J.}~\bibnamefont {Liu}}, \bibinfo
  {author} {\bibfnamefont {J.}~\bibnamefont {Sun}}, \bibinfo {author}
  {\bibfnamefont {J.}~\bibnamefont {Cheng}}, \bibinfo {author} {\bibfnamefont
  {J.}~\bibnamefont {Liu}}, \bibinfo {author} {\bibfnamefont {Y.}~\bibnamefont
  {Yang}}, \bibinfo {author} {\bibfnamefont {R.}~\bibnamefont {Zhang}},
  \bibinfo {author} {\bibfnamefont {L.}~\bibnamefont {Deng}}, \bibinfo {author}
  {\bibfnamefont {W.}~\bibnamefont {Jing}}, \bibinfo {author} {\bibfnamefont
  {Y.}~\bibnamefont {Huang}}, \bibinfo {author} {\bibfnamefont
  {Y.}~\bibnamefont {Shi}}, \bibinfo {author} {\bibfnamefont {M.}~\bibnamefont
  {Ye}}, \bibinfo {author} {\bibfnamefont {S.}~\bibnamefont {Qiao}}, \bibinfo
  {author} {\bibfnamefont {Y.}~\bibnamefont {Wang}}, \bibinfo {author}
  {\bibfnamefont {Y.}~\bibnamefont {Guo}}, \bibinfo {author} {\bibfnamefont
  {D.}~\bibnamefont {Feng}},\ and\ \bibinfo {author} {\bibfnamefont
  {D.}~\bibnamefont {Shen}},\ }\href
  {https://doi.org/10.1103/PhysRevLett.133.206401} {\bibfield  {journal}
  {\bibinfo  {journal} {Phys. Rev. Lett.}\ }\textbf {\bibinfo {volume} {133}},\
  \bibinfo {pages} {206401} (\bibinfo {year} {2024})}\BibitemShut {NoStop}%
\bibitem [{\citenamefont {Duan}\ \emph {et~al.}(2025)\citenamefont {Duan},
  \citenamefont {Zhang}, \citenamefont {Zhu}, \citenamefont {Liu},
  \citenamefont {Zhang}, \citenamefont {\ifmmode \check{Z}\else
  \v{Z}\fi{}uti\ifmmode~\acute{c}\else \'{c}\fi{}},\ and\ \citenamefont
  {Zhou}}]{Duan2025Antiferroelectric}%
  \BibitemOpen
  \bibfield  {author} {\bibinfo {author} {\bibfnamefont {X.}~\bibnamefont
  {Duan}}, \bibinfo {author} {\bibfnamefont {J.}~\bibnamefont {Zhang}},
  \bibinfo {author} {\bibfnamefont {Z.}~\bibnamefont {Zhu}}, \bibinfo {author}
  {\bibfnamefont {Y.}~\bibnamefont {Liu}}, \bibinfo {author} {\bibfnamefont
  {Z.}~\bibnamefont {Zhang}}, \bibinfo {author} {\bibfnamefont
  {I.}~\bibnamefont {\ifmmode \check{Z}\else
  \v{Z}\fi{}uti\ifmmode~\acute{c}\else \'{c}\fi{}}},\ and\ \bibinfo {author}
  {\bibfnamefont {T.}~\bibnamefont {Zhou}},\ }\href
  {https://doi.org/10.1103/PhysRevLett.134.106801} {\bibfield  {journal}
  {\bibinfo  {journal} {Phys. Rev. Lett.}\ }\textbf {\bibinfo {volume} {134}},\
  \bibinfo {pages} {106801} (\bibinfo {year} {2025})}\BibitemShut {NoStop}%
\bibitem [{\citenamefont {Gu}\ \emph {et~al.}(2025)\citenamefont {Gu},
  \citenamefont {Liu}, \citenamefont {Zhu}, \citenamefont {Yananose},
  \citenamefont {Chen}, \citenamefont {Hu}, \citenamefont {Stroppa},\ and\
  \citenamefont {Liu}}]{Gu2025Ferroelectric}%
  \BibitemOpen
  \bibfield  {author} {\bibinfo {author} {\bibfnamefont {M.}~\bibnamefont
  {Gu}}, \bibinfo {author} {\bibfnamefont {Y.}~\bibnamefont {Liu}}, \bibinfo
  {author} {\bibfnamefont {H.}~\bibnamefont {Zhu}}, \bibinfo {author}
  {\bibfnamefont {K.}~\bibnamefont {Yananose}}, \bibinfo {author}
  {\bibfnamefont {X.}~\bibnamefont {Chen}}, \bibinfo {author} {\bibfnamefont
  {Y.}~\bibnamefont {Hu}}, \bibinfo {author} {\bibfnamefont {A.}~\bibnamefont
  {Stroppa}},\ and\ \bibinfo {author} {\bibfnamefont {Q.}~\bibnamefont {Liu}},\
  }\href {https://doi.org/10.1103/PhysRevLett.134.106802} {\bibfield  {journal}
  {\bibinfo  {journal} {Phys. Rev. Lett.}\ }\textbf {\bibinfo {volume} {134}},\
  \bibinfo {pages} {106802} (\bibinfo {year} {2025})}\BibitemShut {NoStop}%
\bibitem [{\citenamefont {Wang}\ \emph {et~al.}(2025)\citenamefont {Wang},
  \citenamefont {Zhang}, \citenamefont {Zhang}, \citenamefont {Sun},
  \citenamefont {Dagotto}, \citenamefont {Xu},\ and\ \citenamefont
  {Hu}}]{Wang2025Spin}%
  \BibitemOpen
  \bibfield  {author} {\bibinfo {author} {\bibfnamefont {Z.-M.}\ \bibnamefont
  {Wang}}, \bibinfo {author} {\bibfnamefont {Y.}~\bibnamefont {Zhang}},
  \bibinfo {author} {\bibfnamefont {S.-B.}\ \bibnamefont {Zhang}}, \bibinfo
  {author} {\bibfnamefont {J.-H.}\ \bibnamefont {Sun}}, \bibinfo {author}
  {\bibfnamefont {E.}~\bibnamefont {Dagotto}}, \bibinfo {author} {\bibfnamefont
  {D.-H.}\ \bibnamefont {Xu}},\ and\ \bibinfo {author} {\bibfnamefont {L.-H.}\
  \bibnamefont {Hu}},\ }\href {https://doi.org/10.1103/cjzw-j4v7} {\bibfield
  {journal} {\bibinfo  {journal} {Phys. Rev. Lett.}\ }\textbf {\bibinfo
  {volume} {135}},\ \bibinfo {pages} {176705} (\bibinfo {year}
  {2025})}\BibitemShut {NoStop}%
\bibitem [{\citenamefont {Zhu}\ \emph {et~al.}(2026{\natexlab{a}})\citenamefont
  {Zhu}, \citenamefont {Huang}, \citenamefont {Chen}, \citenamefont {Cui},
  \citenamefont {Duan}, \citenamefont {Zhang}, \citenamefont {\ifmmode
  \check{Z}\else \v{Z}\fi{}uti\ifmmode~\acute{c}\else \'{c}\fi{}},\ and\
  \citenamefont {Zhou}}]{Zhu2026Altermagnetic}%
  \BibitemOpen
  \bibfield  {author} {\bibinfo {author} {\bibfnamefont {Z.}~\bibnamefont
  {Zhu}}, \bibinfo {author} {\bibfnamefont {R.}~\bibnamefont {Huang}}, \bibinfo
  {author} {\bibfnamefont {X.}~\bibnamefont {Chen}}, \bibinfo {author}
  {\bibfnamefont {Z.}~\bibnamefont {Cui}}, \bibinfo {author} {\bibfnamefont
  {X.}~\bibnamefont {Duan}}, \bibinfo {author} {\bibfnamefont {J.}~\bibnamefont
  {Zhang}}, \bibinfo {author} {\bibfnamefont {I.}~\bibnamefont {\ifmmode
  \check{Z}\else \v{Z}\fi{}uti\ifmmode~\acute{c}\else \'{c}\fi{}}},\ and\
  \bibinfo {author} {\bibfnamefont {T.}~\bibnamefont {Zhou}},\ }\href
  {https://doi.org/10.1103/kqy8-myz1} {\bibfield  {journal} {\bibinfo
  {journal} {Phys. Rev. Lett.}\ }\textbf {\bibinfo {volume} {136}},\ \bibinfo
  {pages} {186702} (\bibinfo {year} {2026}{\natexlab{a}})}\BibitemShut
  {NoStop}%
\bibitem [{\citenamefont {Song}\ \emph
  {et~al.}(2025{\natexlab{a}})\citenamefont {Song}, \citenamefont {Bai},
  \citenamefont {Zhou}, \citenamefont {Han}, \citenamefont {Reichlova},
  \citenamefont {Dil}, \citenamefont {Liu}, \citenamefont {Chen},\ and\
  \citenamefont {Pan}}]{song2025altermagnets}%
  \BibitemOpen
  \bibfield  {author} {\bibinfo {author} {\bibfnamefont {C.}~\bibnamefont
  {Song}}, \bibinfo {author} {\bibfnamefont {H.}~\bibnamefont {Bai}}, \bibinfo
  {author} {\bibfnamefont {Z.}~\bibnamefont {Zhou}}, \bibinfo {author}
  {\bibfnamefont {L.}~\bibnamefont {Han}}, \bibinfo {author} {\bibfnamefont
  {H.}~\bibnamefont {Reichlova}}, \bibinfo {author} {\bibfnamefont {J.~H.}\
  \bibnamefont {Dil}}, \bibinfo {author} {\bibfnamefont {J.}~\bibnamefont
  {Liu}}, \bibinfo {author} {\bibfnamefont {X.}~\bibnamefont {Chen}},\ and\
  \bibinfo {author} {\bibfnamefont {F.}~\bibnamefont {Pan}},\ }\href
  {https://doi.org/10.1038/s41578-025-00779-1} {\bibfield  {journal} {\bibinfo
  {journal} {Nat. Rev. Mater.}\ }\textbf {\bibinfo {volume} {10}},\ \bibinfo
  {pages} {473} (\bibinfo {year} {2025}{\natexlab{a}})}\BibitemShut {NoStop}%
\bibitem [{\citenamefont {Ma}\ \emph {et~al.}(2021)\citenamefont {Ma},
  \citenamefont {Hu}, \citenamefont {Li}, \citenamefont {Liu}, \citenamefont
  {Yao}, \citenamefont {Jia},\ and\ \citenamefont
  {Liu}}]{ma2021multifunctional}%
  \BibitemOpen
  \bibfield  {author} {\bibinfo {author} {\bibfnamefont {H.-Y.}\ \bibnamefont
  {Ma}}, \bibinfo {author} {\bibfnamefont {M.}~\bibnamefont {Hu}}, \bibinfo
  {author} {\bibfnamefont {N.}~\bibnamefont {Li}}, \bibinfo {author}
  {\bibfnamefont {J.}~\bibnamefont {Liu}}, \bibinfo {author} {\bibfnamefont
  {W.}~\bibnamefont {Yao}}, \bibinfo {author} {\bibfnamefont {J.-F.}\
  \bibnamefont {Jia}},\ and\ \bibinfo {author} {\bibfnamefont {J.}~\bibnamefont
  {Liu}},\ }\href {https://doi.org/10.1038/s41467-021-23127-7} {\bibfield
  {journal} {\bibinfo  {journal} {Nat. Commun.}\ }\textbf {\bibinfo {volume}
  {12}},\ \bibinfo {pages} {2846} (\bibinfo {year} {2021})}\BibitemShut
  {NoStop}%
\bibitem [{\citenamefont {Karube}\ \emph {et~al.}(2022)\citenamefont {Karube},
  \citenamefont {Tanaka}, \citenamefont {Sugawara}, \citenamefont {Kadoguchi},
  \citenamefont {Kohda},\ and\ \citenamefont {Nitta}}]{Karube2022Observation}%
  \BibitemOpen
  \bibfield  {author} {\bibinfo {author} {\bibfnamefont {S.}~\bibnamefont
  {Karube}}, \bibinfo {author} {\bibfnamefont {T.}~\bibnamefont {Tanaka}},
  \bibinfo {author} {\bibfnamefont {D.}~\bibnamefont {Sugawara}}, \bibinfo
  {author} {\bibfnamefont {N.}~\bibnamefont {Kadoguchi}}, \bibinfo {author}
  {\bibfnamefont {M.}~\bibnamefont {Kohda}},\ and\ \bibinfo {author}
  {\bibfnamefont {J.}~\bibnamefont {Nitta}},\ }\href
  {https://doi.org/10.1103/PhysRevLett.129.137201} {\bibfield  {journal}
  {\bibinfo  {journal} {Phys. Rev. Lett.}\ }\textbf {\bibinfo {volume} {129}},\
  \bibinfo {pages} {137201} (\bibinfo {year} {2022})}\BibitemShut {NoStop}%
\bibitem [{\citenamefont {Regmi}\ \emph {et~al.}(2025)\citenamefont {Regmi},
  \citenamefont {Bhandari}, \citenamefont {Thapa}, \citenamefont {Hao},
  \citenamefont {Sharma}, \citenamefont {McKenzie}, \citenamefont {Chen},
  \citenamefont {Nayak}, \citenamefont {El~Gazzah}, \citenamefont {M{\'a}rkus}
  \emph {et~al.}}]{regmi2025altermagnetism}%
  \BibitemOpen
  \bibfield  {author} {\bibinfo {author} {\bibfnamefont {R.~B.}\ \bibnamefont
  {Regmi}}, \bibinfo {author} {\bibfnamefont {H.}~\bibnamefont {Bhandari}},
  \bibinfo {author} {\bibfnamefont {B.}~\bibnamefont {Thapa}}, \bibinfo
  {author} {\bibfnamefont {Y.}~\bibnamefont {Hao}}, \bibinfo {author}
  {\bibfnamefont {N.}~\bibnamefont {Sharma}}, \bibinfo {author} {\bibfnamefont
  {J.}~\bibnamefont {McKenzie}}, \bibinfo {author} {\bibfnamefont
  {X.}~\bibnamefont {Chen}}, \bibinfo {author} {\bibfnamefont {A.}~\bibnamefont
  {Nayak}}, \bibinfo {author} {\bibfnamefont {M.}~\bibnamefont {El~Gazzah}},
  \bibinfo {author} {\bibfnamefont {B.~G.}\ \bibnamefont {M{\'a}rkus}}, \emph
  {et~al.},\ }\href {https://doi.org/10.1038/s41467-025-58642-4} {\bibfield
  {journal} {\bibinfo  {journal} {Nat. Commun.}\ }\textbf {\bibinfo {volume}
  {16}},\ \bibinfo {pages} {4399} (\bibinfo {year} {2025})}\BibitemShut
  {NoStop}%
\bibitem [{\citenamefont {Han}\ \emph {et~al.}(2024)\citenamefont {Han},
  \citenamefont {Fu}, \citenamefont {Peng}, \citenamefont {Cheng},
  \citenamefont {Dai}, \citenamefont {Liu}, \citenamefont {Li}, \citenamefont
  {Zhang}, \citenamefont {Zhu}, \citenamefont {Bai} \emph
  {et~al.}}]{han2024electrical}%
  \BibitemOpen
  \bibfield  {author} {\bibinfo {author} {\bibfnamefont {L.}~\bibnamefont
  {Han}}, \bibinfo {author} {\bibfnamefont {X.}~\bibnamefont {Fu}}, \bibinfo
  {author} {\bibfnamefont {R.}~\bibnamefont {Peng}}, \bibinfo {author}
  {\bibfnamefont {X.}~\bibnamefont {Cheng}}, \bibinfo {author} {\bibfnamefont
  {J.}~\bibnamefont {Dai}}, \bibinfo {author} {\bibfnamefont {L.}~\bibnamefont
  {Liu}}, \bibinfo {author} {\bibfnamefont {Y.}~\bibnamefont {Li}}, \bibinfo
  {author} {\bibfnamefont {Y.}~\bibnamefont {Zhang}}, \bibinfo {author}
  {\bibfnamefont {W.}~\bibnamefont {Zhu}}, \bibinfo {author} {\bibfnamefont
  {H.}~\bibnamefont {Bai}}, \emph {et~al.},\ }\href
  {https://doi.org/10.1126/sciadv.adn0479} {\bibfield  {journal} {\bibinfo
  {journal} {Sci. Adv.}\ }\textbf {\bibinfo {volume} {10}},\ \bibinfo {pages}
  {eadn0479} (\bibinfo {year} {2024})}\BibitemShut {NoStop}%
\bibitem [{\citenamefont {Jiang}\ \emph {et~al.}(2025)\citenamefont {Jiang},
  \citenamefont {Hu}, \citenamefont {Bai}, \citenamefont {Song}, \citenamefont
  {Mu}, \citenamefont {Qu}, \citenamefont {Li}, \citenamefont {Zhu},
  \citenamefont {Pi}, \citenamefont {Wei} \emph {et~al.}}]{jiang2025metallic}%
  \BibitemOpen
  \bibfield  {author} {\bibinfo {author} {\bibfnamefont {B.}~\bibnamefont
  {Jiang}}, \bibinfo {author} {\bibfnamefont {M.}~\bibnamefont {Hu}}, \bibinfo
  {author} {\bibfnamefont {J.}~\bibnamefont {Bai}}, \bibinfo {author}
  {\bibfnamefont {Z.}~\bibnamefont {Song}}, \bibinfo {author} {\bibfnamefont
  {C.}~\bibnamefont {Mu}}, \bibinfo {author} {\bibfnamefont {G.}~\bibnamefont
  {Qu}}, \bibinfo {author} {\bibfnamefont {W.}~\bibnamefont {Li}}, \bibinfo
  {author} {\bibfnamefont {W.}~\bibnamefont {Zhu}}, \bibinfo {author}
  {\bibfnamefont {H.}~\bibnamefont {Pi}}, \bibinfo {author} {\bibfnamefont
  {Z.}~\bibnamefont {Wei}}, \emph {et~al.},\ }\href
  {https://doi.org/10.1038/s41567-025-02822-y} {\bibfield  {journal} {\bibinfo
  {journal} {Nat. Phys.}\ }\textbf {\bibinfo {volume} {21}},\ \bibinfo {pages}
  {754} (\bibinfo {year} {2025})}\BibitemShut {NoStop}%
\bibitem [{\citenamefont {Zhang}\ \emph {et~al.}(2025)\citenamefont {Zhang},
  \citenamefont {Cheng}, \citenamefont {Yin}, \citenamefont {Liu},
  \citenamefont {Deng}, \citenamefont {Qiao}, \citenamefont {Shi},
  \citenamefont {Zhang}, \citenamefont {Lin}, \citenamefont {Liu} \emph
  {et~al.}}]{zhang2025crystal}%
  \BibitemOpen
  \bibfield  {author} {\bibinfo {author} {\bibfnamefont {F.}~\bibnamefont
  {Zhang}}, \bibinfo {author} {\bibfnamefont {X.}~\bibnamefont {Cheng}},
  \bibinfo {author} {\bibfnamefont {Z.}~\bibnamefont {Yin}}, \bibinfo {author}
  {\bibfnamefont {C.}~\bibnamefont {Liu}}, \bibinfo {author} {\bibfnamefont
  {L.}~\bibnamefont {Deng}}, \bibinfo {author} {\bibfnamefont {Y.}~\bibnamefont
  {Qiao}}, \bibinfo {author} {\bibfnamefont {Z.}~\bibnamefont {Shi}}, \bibinfo
  {author} {\bibfnamefont {S.}~\bibnamefont {Zhang}}, \bibinfo {author}
  {\bibfnamefont {J.}~\bibnamefont {Lin}}, \bibinfo {author} {\bibfnamefont
  {Z.}~\bibnamefont {Liu}}, \emph {et~al.},\ }\href
  {https://doi.org/10.1038/s41567-025-02864-2} {\bibfield  {journal} {\bibinfo
  {journal} {Nat. Phys.}\ }\textbf {\bibinfo {volume} {21}},\ \bibinfo {pages}
  {760} (\bibinfo {year} {2025})}\BibitemShut {NoStop}%
\bibitem [{\citenamefont {Liu}\ \emph {et~al.}(2024)\citenamefont {Liu},
  \citenamefont {Yu},\ and\ \citenamefont {Liu}}]{Liu2024Twisted}%
  \BibitemOpen
  \bibfield  {author} {\bibinfo {author} {\bibfnamefont {Y.}~\bibnamefont
  {Liu}}, \bibinfo {author} {\bibfnamefont {J.}~\bibnamefont {Yu}},\ and\
  \bibinfo {author} {\bibfnamefont {C.-C.}\ \bibnamefont {Liu}},\ }\href
  {https://doi.org/10.1103/PhysRevLett.133.206702} {\bibfield  {journal}
  {\bibinfo  {journal} {Phys. Rev. Lett.}\ }\textbf {\bibinfo {volume} {133}},\
  \bibinfo {pages} {206702} (\bibinfo {year} {2024})}\BibitemShut {NoStop}%
\bibitem [{\citenamefont {Lin}\ \emph {et~al.}(2025)\citenamefont {Lin},
  \citenamefont {Zhang}, \citenamefont {Lu},\ and\ \citenamefont
  {Xie}}]{Lin2025Coulomb}%
  \BibitemOpen
  \bibfield  {author} {\bibinfo {author} {\bibfnamefont {H.-J.}\ \bibnamefont
  {Lin}}, \bibinfo {author} {\bibfnamefont {S.-B.}\ \bibnamefont {Zhang}},
  \bibinfo {author} {\bibfnamefont {H.-Z.}\ \bibnamefont {Lu}},\ and\ \bibinfo
  {author} {\bibfnamefont {X.~C.}\ \bibnamefont {Xie}},\ }\href
  {https://doi.org/10.1103/PhysRevLett.134.136301} {\bibfield  {journal}
  {\bibinfo  {journal} {Phys. Rev. Lett.}\ }\textbf {\bibinfo {volume} {134}},\
  \bibinfo {pages} {136301} (\bibinfo {year} {2025})}\BibitemShut {NoStop}%
\bibitem [{\citenamefont {Leeb}\ \emph {et~al.}(2024)\citenamefont {Leeb},
  \citenamefont {Mook}, \citenamefont {\ifmmode~\check{S}\else
  \v{S}\fi{}mejkal},\ and\ \citenamefont {Knolle}}]{Leeb2024Spontaneous}%
  \BibitemOpen
  \bibfield  {author} {\bibinfo {author} {\bibfnamefont {V.}~\bibnamefont
  {Leeb}}, \bibinfo {author} {\bibfnamefont {A.}~\bibnamefont {Mook}}, \bibinfo
  {author} {\bibfnamefont {L.}~\bibnamefont {\ifmmode~\check{S}\else
  \v{S}\fi{}mejkal}},\ and\ \bibinfo {author} {\bibfnamefont {J.}~\bibnamefont
  {Knolle}},\ }\href {https://doi.org/10.1103/PhysRevLett.132.236701}
  {\bibfield  {journal} {\bibinfo  {journal} {Phys. Rev. Lett.}\ }\textbf
  {\bibinfo {volume} {132}},\ \bibinfo {pages} {236701} (\bibinfo {year}
  {2024})}\BibitemShut {NoStop}%
\bibitem [{\citenamefont {Wu}\ \emph {et~al.}(2007)\citenamefont {Wu},
  \citenamefont {Sun}, \citenamefont {Fradkin},\ and\ \citenamefont
  {Zhang}}]{Wu2017Fermi}%
  \BibitemOpen
  \bibfield  {author} {\bibinfo {author} {\bibfnamefont {C.}~\bibnamefont
  {Wu}}, \bibinfo {author} {\bibfnamefont {K.}~\bibnamefont {Sun}}, \bibinfo
  {author} {\bibfnamefont {E.}~\bibnamefont {Fradkin}},\ and\ \bibinfo {author}
  {\bibfnamefont {S.-C.}\ \bibnamefont {Zhang}},\ }\href
  {https://doi.org/10.1103/PhysRevB.75.115103} {\bibfield  {journal} {\bibinfo
  {journal} {Phys. Rev. B}\ }\textbf {\bibinfo {volume} {75}},\ \bibinfo
  {pages} {115103} (\bibinfo {year} {2007})}\BibitemShut {NoStop}%
\bibitem [{\citenamefont {Yarmohammadi}\ \emph
  {et~al.}(2026{\natexlab{a}})\citenamefont {Yarmohammadi}, \citenamefont
  {Berritta}, \citenamefont {Bukov}, \citenamefont {\ifmmode~\check{S}\else
  \v{S}\fi{}mejkal}, \citenamefont {Linder},\ and\ \citenamefont
  {Oppeneer}}]{xt23-9pnv}%
  \BibitemOpen
  \bibfield  {author} {\bibinfo {author} {\bibfnamefont {M.}~\bibnamefont
  {Yarmohammadi}}, \bibinfo {author} {\bibfnamefont {M.}~\bibnamefont
  {Berritta}}, \bibinfo {author} {\bibfnamefont {M.}~\bibnamefont {Bukov}},
  \bibinfo {author} {\bibfnamefont {L.}~\bibnamefont {\ifmmode~\check{S}\else
  \v{S}\fi{}mejkal}}, \bibinfo {author} {\bibfnamefont {J.}~\bibnamefont
  {Linder}},\ and\ \bibinfo {author} {\bibfnamefont {P.~M.}\ \bibnamefont
  {Oppeneer}},\ }\href {https://doi.org/10.1103/xt23-9pnv} {\bibfield
  {journal} {\bibinfo  {journal} {Phys. Rev. B}\ }\textbf {\bibinfo {volume}
  {113}},\ \bibinfo {pages} {L060403} (\bibinfo {year}
  {2026}{\natexlab{a}})}\BibitemShut {NoStop}%
\bibitem [{\citenamefont {Fukaya}\ \emph {et~al.}(2025)\citenamefont {Fukaya},
  \citenamefont {Lu}, \citenamefont {Yada}, \citenamefont {Tanaka},\ and\
  \citenamefont {Cayao}}]{fukaya2025superconducting}%
  \BibitemOpen
  \bibfield  {author} {\bibinfo {author} {\bibfnamefont {Y.}~\bibnamefont
  {Fukaya}}, \bibinfo {author} {\bibfnamefont {B.}~\bibnamefont {Lu}}, \bibinfo
  {author} {\bibfnamefont {K.}~\bibnamefont {Yada}}, \bibinfo {author}
  {\bibfnamefont {Y.}~\bibnamefont {Tanaka}},\ and\ \bibinfo {author}
  {\bibfnamefont {J.}~\bibnamefont {Cayao}},\ }\href@noop {} {\bibfield
  {journal} {\bibinfo  {journal} {J. Phys. Condens. Matter}\ }\textbf {\bibinfo
  {volume} {37}},\ \bibinfo {pages} {313003} (\bibinfo {year}
  {2025})}\BibitemShut {NoStop}%
\bibitem [{\citenamefont {Chen}\ \emph {et~al.}(2026)\citenamefont {Chen},
  \citenamefont {Huang}, \citenamefont {Wu}, \citenamefont {Yan},\ and\
  \citenamefont {Wei}}]{chen2026identifying}%
  \BibitemOpen
  \bibfield  {author} {\bibinfo {author} {\bibfnamefont {Y.}~\bibnamefont
  {Chen}}, \bibinfo {author} {\bibfnamefont {Q.}~\bibnamefont {Huang}},
  \bibinfo {author} {\bibfnamefont {Y.}~\bibnamefont {Wu}}, \bibinfo {author}
  {\bibfnamefont {X.}~\bibnamefont {Yan}},\ and\ \bibinfo {author}
  {\bibfnamefont {S.-H.}\ \bibnamefont {Wei}},\ }\href@noop {} {\bibfield
  {journal} {\bibinfo  {journal} {arXiv preprint arXiv:2605.27151}\ } (\bibinfo
  {year} {2026})}\BibitemShut {NoStop}%
\bibitem [{\citenamefont {Song}\ \emph
  {et~al.}(2025{\natexlab{b}})\citenamefont {Song}, \citenamefont
  {Stavri{\'c}}, \citenamefont {Barone}, \citenamefont {Droghetti},
  \citenamefont {Antonenko}, \citenamefont {Venderbos}, \citenamefont
  {Occhialini}, \citenamefont {Ilyas}, \citenamefont {Erge{\c{c}}en},
  \citenamefont {Gedik} \emph {et~al.}}]{song2025electrical}%
  \BibitemOpen
  \bibfield  {author} {\bibinfo {author} {\bibfnamefont {Q.}~\bibnamefont
  {Song}}, \bibinfo {author} {\bibfnamefont {S.}~\bibnamefont {Stavri{\'c}}},
  \bibinfo {author} {\bibfnamefont {P.}~\bibnamefont {Barone}}, \bibinfo
  {author} {\bibfnamefont {A.}~\bibnamefont {Droghetti}}, \bibinfo {author}
  {\bibfnamefont {D.~S.}\ \bibnamefont {Antonenko}}, \bibinfo {author}
  {\bibfnamefont {J.~W.}\ \bibnamefont {Venderbos}}, \bibinfo {author}
  {\bibfnamefont {C.~A.}\ \bibnamefont {Occhialini}}, \bibinfo {author}
  {\bibfnamefont {B.}~\bibnamefont {Ilyas}}, \bibinfo {author} {\bibfnamefont
  {E.}~\bibnamefont {Erge{\c{c}}en}}, \bibinfo {author} {\bibfnamefont
  {N.}~\bibnamefont {Gedik}}, \emph {et~al.},\ }\href
  {https://doi.org/10.1038/s41586-025-09034-7} {\bibfield  {journal} {\bibinfo
  {journal} {Nature}\ }\textbf {\bibinfo {volume} {642}},\ \bibinfo {pages}
  {64} (\bibinfo {year} {2025}{\natexlab{b}})}\BibitemShut {NoStop}%
\bibitem [{\citenamefont {Yamada}\ \emph {et~al.}(2025)\citenamefont {Yamada},
  \citenamefont {Birch}, \citenamefont {Baral}, \citenamefont {Okumura},
  \citenamefont {Nakano}, \citenamefont {Gao}, \citenamefont {Ezawa},
  \citenamefont {Nomoto}, \citenamefont {Masell}, \citenamefont {Ishihara}
  \emph {et~al.}}]{yamada2025metallic}%
  \BibitemOpen
  \bibfield  {author} {\bibinfo {author} {\bibfnamefont {R.}~\bibnamefont
  {Yamada}}, \bibinfo {author} {\bibfnamefont {M.~T.}\ \bibnamefont {Birch}},
  \bibinfo {author} {\bibfnamefont {P.~R.}\ \bibnamefont {Baral}}, \bibinfo
  {author} {\bibfnamefont {S.}~\bibnamefont {Okumura}}, \bibinfo {author}
  {\bibfnamefont {R.}~\bibnamefont {Nakano}}, \bibinfo {author} {\bibfnamefont
  {S.}~\bibnamefont {Gao}}, \bibinfo {author} {\bibfnamefont {M.}~\bibnamefont
  {Ezawa}}, \bibinfo {author} {\bibfnamefont {T.}~\bibnamefont {Nomoto}},
  \bibinfo {author} {\bibfnamefont {J.}~\bibnamefont {Masell}}, \bibinfo
  {author} {\bibfnamefont {Y.}~\bibnamefont {Ishihara}}, \emph {et~al.},\
  }\href {https://doi.org/10.1038/s41586-025-09633-4} {\bibfield  {journal}
  {\bibinfo  {journal} {Nature}\ }\textbf {\bibinfo {volume} {646}},\ \bibinfo
  {pages} {837} (\bibinfo {year} {2025})}\BibitemShut {NoStop}%
\bibitem [{\citenamefont {Yu}\ \emph {et~al.}(2025)\citenamefont {Yu},
  \citenamefont {Lyngby}, \citenamefont {Shishidou}, \citenamefont {Roig},
  \citenamefont {Kreisel}, \citenamefont {Weinert}, \citenamefont {Andersen},\
  and\ \citenamefont {Agterberg}}]{Yu2025Odd}%
  \BibitemOpen
  \bibfield  {author} {\bibinfo {author} {\bibfnamefont {Y.}~\bibnamefont
  {Yu}}, \bibinfo {author} {\bibfnamefont {M.~B.}\ \bibnamefont {Lyngby}},
  \bibinfo {author} {\bibfnamefont {T.}~\bibnamefont {Shishidou}}, \bibinfo
  {author} {\bibfnamefont {M.}~\bibnamefont {Roig}}, \bibinfo {author}
  {\bibfnamefont {A.}~\bibnamefont {Kreisel}}, \bibinfo {author} {\bibfnamefont
  {M.}~\bibnamefont {Weinert}}, \bibinfo {author} {\bibfnamefont {B.~M.}\
  \bibnamefont {Andersen}},\ and\ \bibinfo {author} {\bibfnamefont {D.~F.}\
  \bibnamefont {Agterberg}},\ }\href {https://doi.org/10.1103/zk69-k6b2}
  {\bibfield  {journal} {\bibinfo  {journal} {Phys. Rev. Lett.}\ }\textbf
  {\bibinfo {volume} {135}},\ \bibinfo {pages} {046701} (\bibinfo {year}
  {2025})}\BibitemShut {NoStop}%
\bibitem [{\citenamefont {Hellenes}\ \emph {et~al.}(2023)\citenamefont
  {Hellenes}, \citenamefont {Jungwirth}, \citenamefont {Jaeschke-Ubiergo},
  \citenamefont {Chakraborty}, \citenamefont {Sinova},\ and\ \citenamefont
  {{\v{S}}mejkal}}]{hellenes2023p}%
  \BibitemOpen
  \bibfield  {author} {\bibinfo {author} {\bibfnamefont {A.~B.}\ \bibnamefont
  {Hellenes}}, \bibinfo {author} {\bibfnamefont {T.}~\bibnamefont {Jungwirth}},
  \bibinfo {author} {\bibfnamefont {R.}~\bibnamefont {Jaeschke-Ubiergo}},
  \bibinfo {author} {\bibfnamefont {A.}~\bibnamefont {Chakraborty}}, \bibinfo
  {author} {\bibfnamefont {J.}~\bibnamefont {Sinova}},\ and\ \bibinfo {author}
  {\bibfnamefont {L.}~\bibnamefont {{\v{S}}mejkal}},\ }\href@noop {} {\bibfield
   {journal} {\bibinfo  {journal} {arXiv preprint arXiv:2309.01607}\ }
  (\bibinfo {year} {2023})}\BibitemShut {NoStop}%
\bibitem [{\citenamefont {Zhang}\ \emph {et~al.}(2026)\citenamefont {Zhang},
  \citenamefont {Jiang}, \citenamefont {Shen}, \citenamefont {Yuan},
  \citenamefont {Yoo}, \citenamefont {Ma}, \citenamefont {Ye}, \citenamefont
  {Liu}, \citenamefont {Liu}, \citenamefont {Kim} \emph
  {et~al.}}]{zhang2026quenching}%
  \BibitemOpen
  \bibfield  {author} {\bibinfo {author} {\bibfnamefont {X.}~\bibnamefont
  {Zhang}}, \bibinfo {author} {\bibfnamefont {Z.}~\bibnamefont {Jiang}},
  \bibinfo {author} {\bibfnamefont {S.}~\bibnamefont {Shen}}, \bibinfo {author}
  {\bibfnamefont {J.}~\bibnamefont {Yuan}}, \bibinfo {author} {\bibfnamefont
  {J.}~\bibnamefont {Yoo}}, \bibinfo {author} {\bibfnamefont {X.}~\bibnamefont
  {Ma}}, \bibinfo {author} {\bibfnamefont {M.}~\bibnamefont {Ye}}, \bibinfo
  {author} {\bibfnamefont {J.}~\bibnamefont {Liu}}, \bibinfo {author}
  {\bibfnamefont {Z.}~\bibnamefont {Liu}}, \bibinfo {author} {\bibfnamefont
  {C.}~\bibnamefont {Kim}}, \emph {et~al.},\ }\href@noop {} {\bibfield
  {journal} {\bibinfo  {journal} {arXiv preprint arXiv:2606.02420}\ } (\bibinfo
  {year} {2026})}\BibitemShut {NoStop}%
\bibitem [{\citenamefont {Hayami}\ \emph {et~al.}(2020)\citenamefont {Hayami},
  \citenamefont {Yanagi},\ and\ \citenamefont
  {Kusunose}}]{Hayami2020Spontaneous}%
  \BibitemOpen
  \bibfield  {author} {\bibinfo {author} {\bibfnamefont {S.}~\bibnamefont
  {Hayami}}, \bibinfo {author} {\bibfnamefont {Y.}~\bibnamefont {Yanagi}},\
  and\ \bibinfo {author} {\bibfnamefont {H.}~\bibnamefont {Kusunose}},\ }\href
  {https://doi.org/10.1103/PhysRevB.101.220403} {\bibfield  {journal} {\bibinfo
   {journal} {Phys. Rev. B}\ }\textbf {\bibinfo {volume} {101}},\ \bibinfo
  {pages} {220403(R)} (\bibinfo {year} {2020})}\BibitemShut {NoStop}%
\bibitem [{\citenamefont {Li}\ and\ \citenamefont {Sukhachov}(2026)}]{li2026p}%
  \BibitemOpen
  \bibfield  {author} {\bibinfo {author} {\bibfnamefont {Y.}~\bibnamefont
  {Li}}\ and\ \bibinfo {author} {\bibfnamefont {P.}~\bibnamefont {Sukhachov}},\
  }\href@noop {} {\bibfield  {journal} {\bibinfo  {journal} {arXiv preprint
  arXiv:2604.18695}\ } (\bibinfo {year} {2026})}\BibitemShut {NoStop}%
\bibitem [{\citenamefont {Zhuang}\ \emph {et~al.}(2025)\citenamefont {Zhuang},
  \citenamefont {Zhu}, \citenamefont {Liu}, \citenamefont {Wu},\ and\
  \citenamefont {Yan}}]{zhuang2025odd}%
  \BibitemOpen
  \bibfield  {author} {\bibinfo {author} {\bibfnamefont {Z.-Y.}\ \bibnamefont
  {Zhuang}}, \bibinfo {author} {\bibfnamefont {D.}~\bibnamefont {Zhu}},
  \bibinfo {author} {\bibfnamefont {D.}~\bibnamefont {Liu}}, \bibinfo {author}
  {\bibfnamefont {Z.}~\bibnamefont {Wu}},\ and\ \bibinfo {author}
  {\bibfnamefont {Z.}~\bibnamefont {Yan}},\ }\href@noop {} {\bibfield
  {journal} {\bibinfo  {journal} {arXiv preprint arXiv:2508.18361}\ } (\bibinfo
  {year} {2025})}\BibitemShut {NoStop}%
\bibitem [{\citenamefont {Lin}\ and\ \citenamefont {Vila}(2025)}]{lin2025odd}%
  \BibitemOpen
  \bibfield  {author} {\bibinfo {author} {\bibfnamefont {Y.-P.}\ \bibnamefont
  {Lin}}\ and\ \bibinfo {author} {\bibfnamefont {M.}~\bibnamefont {Vila}},\
  }\href@noop {} {\bibfield  {journal} {\bibinfo  {journal} {arXiv preprint
  arXiv:2503.09602}\ } (\bibinfo {year} {2025})}\BibitemShut {NoStop}%
\bibitem [{\citenamefont {Leeb}\ and\ \citenamefont
  {Knolle}(2026)}]{leeb2026collinear}%
  \BibitemOpen
  \bibfield  {author} {\bibinfo {author} {\bibfnamefont {V.}~\bibnamefont
  {Leeb}}\ and\ \bibinfo {author} {\bibfnamefont {J.}~\bibnamefont {Knolle}},\
  }\href@noop {} {\bibfield  {journal} {\bibinfo  {journal} {arXiv preprint
  arXiv:2601.07418}\ } (\bibinfo {year} {2026})}\BibitemShut {NoStop}%
\bibitem [{\citenamefont {Zhu}\ \emph {et~al.}(2026{\natexlab{b}})\citenamefont
  {Zhu}, \citenamefont {Zhou}, \citenamefont {Wang}, \citenamefont {Wei},\ and\
  \citenamefont {Ruan}}]{Zhu2026Floquet}%
  \BibitemOpen
  \bibfield  {author} {\bibinfo {author} {\bibfnamefont {T.}~\bibnamefont
  {Zhu}}, \bibinfo {author} {\bibfnamefont {D.}~\bibnamefont {Zhou}}, \bibinfo
  {author} {\bibfnamefont {H.}~\bibnamefont {Wang}}, \bibinfo {author}
  {\bibfnamefont {S.-H.}\ \bibnamefont {Wei}},\ and\ \bibinfo {author}
  {\bibfnamefont {J.}~\bibnamefont {Ruan}},\ }\href
  {https://doi.org/10.1103/7ywb-ml2q} {\bibfield  {journal} {\bibinfo
  {journal} {Phys. Rev. Lett.}\ }\textbf {\bibinfo {volume} {136}},\ \bibinfo
  {pages} {126704} (\bibinfo {year} {2026}{\natexlab{b}})}\BibitemShut
  {NoStop}%
\bibitem [{\citenamefont {Huang}\ \emph {et~al.}(2026)\citenamefont {Huang},
  \citenamefont {Qin}, \citenamefont {Zhan}, \citenamefont {Xu}, \citenamefont
  {Ma},\ and\ \citenamefont {Wang}}]{Huang2026Light}%
  \BibitemOpen
  \bibfield  {author} {\bibinfo {author} {\bibfnamefont {S.}~\bibnamefont
  {Huang}}, \bibinfo {author} {\bibfnamefont {Z.}~\bibnamefont {Qin}}, \bibinfo
  {author} {\bibfnamefont {F.}~\bibnamefont {Zhan}}, \bibinfo {author}
  {\bibfnamefont {D.-H.}\ \bibnamefont {Xu}}, \bibinfo {author} {\bibfnamefont
  {D.-S.}\ \bibnamefont {Ma}},\ and\ \bibinfo {author} {\bibfnamefont
  {R.}~\bibnamefont {Wang}},\ }\href {https://doi.org/10.1103/9346-9jpf}
  {\bibfield  {journal} {\bibinfo  {journal} {Phys. Rev. Lett.}\ }\textbf
  {\bibinfo {volume} {136}},\ \bibinfo {pages} {126703} (\bibinfo {year}
  {2026})}\BibitemShut {NoStop}%
\bibitem [{\citenamefont {Liu}\ \emph {et~al.}(2026{\natexlab{a}})\citenamefont
  {Liu}, \citenamefont {Zhuang}, \citenamefont {Zhu}, \citenamefont {Wu},\ and\
  \citenamefont {Yan}}]{Liu2026Light}%
  \BibitemOpen
  \bibfield  {author} {\bibinfo {author} {\bibfnamefont {D.}~\bibnamefont
  {Liu}}, \bibinfo {author} {\bibfnamefont {Z.-Y.}\ \bibnamefont {Zhuang}},
  \bibinfo {author} {\bibfnamefont {D.}~\bibnamefont {Zhu}}, \bibinfo {author}
  {\bibfnamefont {Z.}~\bibnamefont {Wu}},\ and\ \bibinfo {author}
  {\bibfnamefont {Z.}~\bibnamefont {Yan}},\ }\href
  {https://doi.org/10.1103/wnqs-3djt} {\bibfield  {journal} {\bibinfo
  {journal} {Phys. Rev. B}\ }\textbf {\bibinfo {volume} {113}},\ \bibinfo
  {pages} {L060409} (\bibinfo {year} {2026}{\natexlab{a}})}\BibitemShut
  {NoStop}%
\bibitem [{\citenamefont {Li}\ \emph {et~al.}(2026)\citenamefont {Li},
  \citenamefont {Shao},\ and\ \citenamefont {Kovalev}}]{Li2026Floquet}%
  \BibitemOpen
  \bibfield  {author} {\bibinfo {author} {\bibfnamefont {B.}~\bibnamefont
  {Li}}, \bibinfo {author} {\bibfnamefont {D.-F.}\ \bibnamefont {Shao}},\ and\
  \bibinfo {author} {\bibfnamefont {A.~A.}\ \bibnamefont {Kovalev}},\ }\href
  {https://doi.org/10.1103/xzm1-l6yf} {\bibfield  {journal} {\bibinfo
  {journal} {Phys. Rev. Lett.}\ }\textbf {\bibinfo {volume} {136}},\ \bibinfo
  {pages} {166701} (\bibinfo {year} {2026})}\BibitemShut {NoStop}%
\bibitem [{\citenamefont {Tian}\ \emph {et~al.}(2026)\citenamefont {Tian},
  \citenamefont {Zhao}, \citenamefont {Wang}, \citenamefont {Zhang},
  \citenamefont {Kong},\ and\ \citenamefont {Gong}}]{tian2026optically}%
  \BibitemOpen
  \bibfield  {author} {\bibinfo {author} {\bibfnamefont {Y.}~\bibnamefont
  {Tian}}, \bibinfo {author} {\bibfnamefont {C.-H.}\ \bibnamefont {Zhao}},
  \bibinfo {author} {\bibfnamefont {C.-B.}\ \bibnamefont {Wang}}, \bibinfo
  {author} {\bibfnamefont {B.}~\bibnamefont {Zhang}}, \bibinfo {author}
  {\bibfnamefont {X.}~\bibnamefont {Kong}},\ and\ \bibinfo {author}
  {\bibfnamefont {W.-J.}\ \bibnamefont {Gong}},\ }\href@noop {} {\bibfield
  {journal} {\bibinfo  {journal} {arXiv preprint arXiv:2603.11483}\ } (\bibinfo
  {year} {2026})}\BibitemShut {NoStop}%
\bibitem [{\citenamefont {Li}\ \emph {et~al.}(2025)\citenamefont {Li},
  \citenamefont {Li}, \citenamefont {Guan},\ and\ \citenamefont
  {Meng}}]{li2025robust}%
  \BibitemOpen
  \bibfield  {author} {\bibinfo {author} {\bibfnamefont {Z.}~\bibnamefont
  {Li}}, \bibinfo {author} {\bibfnamefont {L.}~\bibnamefont {Li}}, \bibinfo
  {author} {\bibfnamefont {M.}~\bibnamefont {Guan}},\ and\ \bibinfo {author}
  {\bibfnamefont {S.}~\bibnamefont {Meng}},\ }\href@noop {} {\bibfield
  {journal} {\bibinfo  {journal} {arXiv preprint arXiv:2512.06416}\ } (\bibinfo
  {year} {2025})}\BibitemShut {NoStop}%
\bibitem [{\citenamefont {Chakraborty}\ \emph {et~al.}(2025)\citenamefont
  {Chakraborty}, \citenamefont {Birk~Hellenes}, \citenamefont
  {Jaeschke-Ubiergo}, \citenamefont {Jungwirth}, \citenamefont
  {{\v{S}}mejkal},\ and\ \citenamefont {Sinova}}]{chakraborty2025highly}%
  \BibitemOpen
  \bibfield  {author} {\bibinfo {author} {\bibfnamefont {A.}~\bibnamefont
  {Chakraborty}}, \bibinfo {author} {\bibfnamefont {A.}~\bibnamefont
  {Birk~Hellenes}}, \bibinfo {author} {\bibfnamefont {R.}~\bibnamefont
  {Jaeschke-Ubiergo}}, \bibinfo {author} {\bibfnamefont {T.}~\bibnamefont
  {Jungwirth}}, \bibinfo {author} {\bibfnamefont {L.}~\bibnamefont
  {{\v{S}}mejkal}},\ and\ \bibinfo {author} {\bibfnamefont {J.}~\bibnamefont
  {Sinova}},\ }\href {https://doi.org/10.1038/s41467-025-62516-0} {\bibfield
  {journal} {\bibinfo  {journal} {Nat. Commun.}\ }\textbf {\bibinfo {volume}
  {16}},\ \bibinfo {pages} {7270} (\bibinfo {year} {2025})}\BibitemShut
  {NoStop}%
\bibitem [{\citenamefont {Brekke}\ \emph {et~al.}(2024)\citenamefont {Brekke},
  \citenamefont {Sukhachov}, \citenamefont {Giil}, \citenamefont {Brataas},\
  and\ \citenamefont {Linder}}]{Brekke2024}%
  \BibitemOpen
  \bibfield  {author} {\bibinfo {author} {\bibfnamefont {B.}~\bibnamefont
  {Brekke}}, \bibinfo {author} {\bibfnamefont {P.}~\bibnamefont {Sukhachov}},
  \bibinfo {author} {\bibfnamefont {H.~G.}\ \bibnamefont {Giil}}, \bibinfo
  {author} {\bibfnamefont {A.}~\bibnamefont {Brataas}},\ and\ \bibinfo {author}
  {\bibfnamefont {J.}~\bibnamefont {Linder}},\ }\href
  {https://doi.org/10.1103/PhysRevLett.133.236703} {\bibfield  {journal}
  {\bibinfo  {journal} {Phys. Rev. Lett.}\ }\textbf {\bibinfo {volume} {133}},\
  \bibinfo {pages} {236703} (\bibinfo {year} {2024})}\BibitemShut {NoStop}%
\bibitem [{\citenamefont {Luo}\ \emph {et~al.}(2026)\citenamefont {Luo},
  \citenamefont {Sun}, \citenamefont {Feng}, \citenamefont {Tian},\ and\
  \citenamefont {Law}}]{Luo2026}%
  \BibitemOpen
  \bibfield  {author} {\bibinfo {author} {\bibfnamefont {X.-J.}\ \bibnamefont
  {Luo}}, \bibinfo {author} {\bibfnamefont {Z.-T.}\ \bibnamefont {Sun}},
  \bibinfo {author} {\bibfnamefont {X.}~\bibnamefont {Feng}}, \bibinfo {author}
  {\bibfnamefont {M.}~\bibnamefont {Tian}},\ and\ \bibinfo {author}
  {\bibfnamefont {K.~T.}\ \bibnamefont {Law}},\ }\href
  {https://doi.org/10.1103/j9m4-386b} {\bibfield  {journal} {\bibinfo
  {journal} {Phys. Rev. Lett.}\ }\textbf {\bibinfo {volume} {137}},\ \bibinfo
  {pages} {136003} (\bibinfo {year} {2026})}\BibitemShut {NoStop}%
\bibitem [{\citenamefont {Khodas}\ \emph {et~al.}(2026)\citenamefont {Khodas},
  \citenamefont {{\v{S}}mejkal},\ and\ \citenamefont
  {Mazin}}]{khodas2026nonrelativistic}%
  \BibitemOpen
  \bibfield  {author} {\bibinfo {author} {\bibfnamefont {M.}~\bibnamefont
  {Khodas}}, \bibinfo {author} {\bibfnamefont {L.}~\bibnamefont
  {{\v{S}}mejkal}},\ and\ \bibinfo {author} {\bibfnamefont {I.}~\bibnamefont
  {Mazin}},\ }\href {https://doi.org/10.1103/p21c-rtv7} {\bibfield  {journal}
  {\bibinfo  {journal} {Phys. Rev. B}\ }\textbf {\bibinfo {volume} {114}},\
  \bibinfo {pages} {L080505} (\bibinfo {year} {2026})}\BibitemShut {NoStop}%
\bibitem [{\citenamefont {Zhuang}\ \emph {et~al.}(2026)\citenamefont {Zhuang},
  \citenamefont {Hu}, \citenamefont {Zhang}, \citenamefont {Hu},\ and\
  \citenamefont {Yan}}]{zhuang2026mixed}%
  \BibitemOpen
  \bibfield  {author} {\bibinfo {author} {\bibfnamefont {Z.-Y.}\ \bibnamefont
  {Zhuang}}, \bibinfo {author} {\bibfnamefont {J.-X.}\ \bibnamefont {Hu}},
  \bibinfo {author} {\bibfnamefont {S.-B.}\ \bibnamefont {Zhang}}, \bibinfo
  {author} {\bibfnamefont {L.-H.}\ \bibnamefont {Hu}},\ and\ \bibinfo {author}
  {\bibfnamefont {Z.}~\bibnamefont {Yan}},\ }\href@noop {} {\bibfield
  {journal} {\bibinfo  {journal} {arXiv preprint arXiv:2605.05205}\ } (\bibinfo
  {year} {2026})}\BibitemShut {NoStop}%
\bibitem [{\citenamefont {Wang}\ \emph {et~al.}(2013)\citenamefont {Wang},
  \citenamefont {Steinberg}, \citenamefont {Jarillo-Herrero},\ and\
  \citenamefont {Gedik}}]{wang2013observation}%
  \BibitemOpen
  \bibfield  {author} {\bibinfo {author} {\bibfnamefont {Y.}~\bibnamefont
  {Wang}}, \bibinfo {author} {\bibfnamefont {H.}~\bibnamefont {Steinberg}},
  \bibinfo {author} {\bibfnamefont {P.}~\bibnamefont {Jarillo-Herrero}},\ and\
  \bibinfo {author} {\bibfnamefont {N.}~\bibnamefont {Gedik}},\ }\href
  {https://doi.org/10.1126/science.1239834} {\bibfield  {journal} {\bibinfo
  {journal} {Science}\ }\textbf {\bibinfo {volume} {342}},\ \bibinfo {pages}
  {453} (\bibinfo {year} {2013})}\BibitemShut {NoStop}%
\bibitem [{\citenamefont {Zhou}\ \emph {et~al.}(2023)\citenamefont {Zhou},
  \citenamefont {Bao}, \citenamefont {Fan}, \citenamefont {Zhou}, \citenamefont
  {Gao}, \citenamefont {Zhong}, \citenamefont {Lin}, \citenamefont {Liu},
  \citenamefont {Yu}, \citenamefont {Tang} \emph
  {et~al.}}]{zhou2023pseudospin}%
  \BibitemOpen
  \bibfield  {author} {\bibinfo {author} {\bibfnamefont {S.}~\bibnamefont
  {Zhou}}, \bibinfo {author} {\bibfnamefont {C.}~\bibnamefont {Bao}}, \bibinfo
  {author} {\bibfnamefont {B.}~\bibnamefont {Fan}}, \bibinfo {author}
  {\bibfnamefont {H.}~\bibnamefont {Zhou}}, \bibinfo {author} {\bibfnamefont
  {Q.}~\bibnamefont {Gao}}, \bibinfo {author} {\bibfnamefont {H.}~\bibnamefont
  {Zhong}}, \bibinfo {author} {\bibfnamefont {T.}~\bibnamefont {Lin}}, \bibinfo
  {author} {\bibfnamefont {H.}~\bibnamefont {Liu}}, \bibinfo {author}
  {\bibfnamefont {P.}~\bibnamefont {Yu}}, \bibinfo {author} {\bibfnamefont
  {P.}~\bibnamefont {Tang}}, \emph {et~al.},\ }\href
  {https://doi.org/10.1038/s41586-022-05610-3} {\bibfield  {journal} {\bibinfo
  {journal} {Nature}\ }\textbf {\bibinfo {volume} {614}},\ \bibinfo {pages}
  {75} (\bibinfo {year} {2023})}\BibitemShut {NoStop}%
\bibitem [{\citenamefont {McIver}\ \emph {et~al.}(2020)\citenamefont {McIver},
  \citenamefont {Schulte}, \citenamefont {Stein}, \citenamefont {Matsuyama},
  \citenamefont {Jotzu}, \citenamefont {Meier},\ and\ \citenamefont
  {Cavalleri}}]{mciver2020light}%
  \BibitemOpen
  \bibfield  {author} {\bibinfo {author} {\bibfnamefont {J.~W.}\ \bibnamefont
  {McIver}}, \bibinfo {author} {\bibfnamefont {B.}~\bibnamefont {Schulte}},
  \bibinfo {author} {\bibfnamefont {F.-U.}\ \bibnamefont {Stein}}, \bibinfo
  {author} {\bibfnamefont {T.}~\bibnamefont {Matsuyama}}, \bibinfo {author}
  {\bibfnamefont {G.}~\bibnamefont {Jotzu}}, \bibinfo {author} {\bibfnamefont
  {G.}~\bibnamefont {Meier}},\ and\ \bibinfo {author} {\bibfnamefont
  {A.}~\bibnamefont {Cavalleri}},\ }\href
  {https://doi.org/10.1038/s41567-019-0698-y} {\bibfield  {journal} {\bibinfo
  {journal} {Nat. Phys.}\ }\textbf {\bibinfo {volume} {16}},\ \bibinfo {pages}
  {38} (\bibinfo {year} {2020})}\BibitemShut {NoStop}%
\bibitem [{\citenamefont {H{\"u}bener}\ \emph {et~al.}(2017)\citenamefont
  {H{\"u}bener}, \citenamefont {Sentef}, \citenamefont {De~Giovannini},
  \citenamefont {Kemper},\ and\ \citenamefont {Rubio}}]{hubener2017creating}%
  \BibitemOpen
  \bibfield  {author} {\bibinfo {author} {\bibfnamefont {H.}~\bibnamefont
  {H{\"u}bener}}, \bibinfo {author} {\bibfnamefont {M.~A.}\ \bibnamefont
  {Sentef}}, \bibinfo {author} {\bibfnamefont {U.}~\bibnamefont
  {De~Giovannini}}, \bibinfo {author} {\bibfnamefont {A.~F.}\ \bibnamefont
  {Kemper}},\ and\ \bibinfo {author} {\bibfnamefont {A.}~\bibnamefont
  {Rubio}},\ }\href {https://doi.org/10.1038/ncomms13940} {\bibfield  {journal}
  {\bibinfo  {journal} {Nat. Commun.}\ }\textbf {\bibinfo {volume} {8}},\
  \bibinfo {pages} {13940} (\bibinfo {year} {2017})}\BibitemShut {NoStop}%
\bibitem [{\citenamefont {Choi}\ \emph {et~al.}(2025)\citenamefont {Choi},
  \citenamefont {Mogi}, \citenamefont {De~Giovannini}, \citenamefont {Azoury},
  \citenamefont {Lv}, \citenamefont {Su}, \citenamefont {H{\"u}bener},
  \citenamefont {Rubio},\ and\ \citenamefont {Gedik}}]{choi2025observation}%
  \BibitemOpen
  \bibfield  {author} {\bibinfo {author} {\bibfnamefont {D.}~\bibnamefont
  {Choi}}, \bibinfo {author} {\bibfnamefont {M.}~\bibnamefont {Mogi}}, \bibinfo
  {author} {\bibfnamefont {U.}~\bibnamefont {De~Giovannini}}, \bibinfo {author}
  {\bibfnamefont {D.}~\bibnamefont {Azoury}}, \bibinfo {author} {\bibfnamefont
  {B.}~\bibnamefont {Lv}}, \bibinfo {author} {\bibfnamefont {Y.}~\bibnamefont
  {Su}}, \bibinfo {author} {\bibfnamefont {H.}~\bibnamefont {H{\"u}bener}},
  \bibinfo {author} {\bibfnamefont {A.}~\bibnamefont {Rubio}},\ and\ \bibinfo
  {author} {\bibfnamefont {N.}~\bibnamefont {Gedik}},\ }\href
  {https://doi.org/10.1038/s41567-025-02888-8} {\bibfield  {journal} {\bibinfo
  {journal} {Nat. Phys.}\ }\textbf {\bibinfo {volume} {21}},\ \bibinfo {pages}
  {1100} (\bibinfo {year} {2025})}\BibitemShut {NoStop}%
\bibitem [{\citenamefont {Bao}\ \emph {et~al.}(2024)\citenamefont {Bao},
  \citenamefont {Sch{\"u}ler}, \citenamefont {Xiao}, \citenamefont {Wang},
  \citenamefont {Zhong}, \citenamefont {Lin}, \citenamefont {Cai},
  \citenamefont {Sheng}, \citenamefont {Tang}, \citenamefont {Zhang} \emph
  {et~al.}}]{bao2024manipulating}%
  \BibitemOpen
  \bibfield  {author} {\bibinfo {author} {\bibfnamefont {C.}~\bibnamefont
  {Bao}}, \bibinfo {author} {\bibfnamefont {M.}~\bibnamefont {Sch{\"u}ler}},
  \bibinfo {author} {\bibfnamefont {T.}~\bibnamefont {Xiao}}, \bibinfo {author}
  {\bibfnamefont {F.}~\bibnamefont {Wang}}, \bibinfo {author} {\bibfnamefont
  {H.}~\bibnamefont {Zhong}}, \bibinfo {author} {\bibfnamefont
  {T.}~\bibnamefont {Lin}}, \bibinfo {author} {\bibfnamefont {X.}~\bibnamefont
  {Cai}}, \bibinfo {author} {\bibfnamefont {T.}~\bibnamefont {Sheng}}, \bibinfo
  {author} {\bibfnamefont {X.}~\bibnamefont {Tang}}, \bibinfo {author}
  {\bibfnamefont {H.}~\bibnamefont {Zhang}}, \emph {et~al.},\ }\href
  {https://doi.org/10.1038/s41467-024-54760-7} {\bibfield  {journal} {\bibinfo
  {journal} {Nat. Commun.}\ }\textbf {\bibinfo {volume} {15}},\ \bibinfo
  {pages} {10535} (\bibinfo {year} {2024})}\BibitemShut {NoStop}%
\bibitem [{\citenamefont {Liu}\ \emph {et~al.}(2025)\citenamefont {Liu},
  \citenamefont {Yang}, \citenamefont {Gaertner}, \citenamefont {Huckabee},
  \citenamefont {Suslov}, \citenamefont {Refael}, \citenamefont {Nathan},
  \citenamefont {Lewandowski}, \citenamefont {Foa~Torres}, \citenamefont {Esin}
  \emph {et~al.}}]{liu2025signatures}%
  \BibitemOpen
  \bibfield  {author} {\bibinfo {author} {\bibfnamefont {Y.}~\bibnamefont
  {Liu}}, \bibinfo {author} {\bibfnamefont {C.}~\bibnamefont {Yang}}, \bibinfo
  {author} {\bibfnamefont {G.}~\bibnamefont {Gaertner}}, \bibinfo {author}
  {\bibfnamefont {J.}~\bibnamefont {Huckabee}}, \bibinfo {author}
  {\bibfnamefont {A.~V.}\ \bibnamefont {Suslov}}, \bibinfo {author}
  {\bibfnamefont {G.}~\bibnamefont {Refael}}, \bibinfo {author} {\bibfnamefont
  {F.}~\bibnamefont {Nathan}}, \bibinfo {author} {\bibfnamefont
  {C.}~\bibnamefont {Lewandowski}}, \bibinfo {author} {\bibfnamefont {L.~E.}\
  \bibnamefont {Foa~Torres}}, \bibinfo {author} {\bibfnamefont
  {I.}~\bibnamefont {Esin}}, \emph {et~al.},\ }\href
  {https://doi.org/10.1038/s41467-025-57335-2} {\bibfield  {journal} {\bibinfo
  {journal} {Nat. Commun.}\ }\textbf {\bibinfo {volume} {16}},\ \bibinfo
  {pages} {2057} (\bibinfo {year} {2025})}\BibitemShut {NoStop}%
\bibitem [{\citenamefont {Merboldt}\ \emph {et~al.}(2025)\citenamefont
  {Merboldt}, \citenamefont {Sch{\"u}ler}, \citenamefont {Schmitt},
  \citenamefont {Bange}, \citenamefont {Bennecke}, \citenamefont {Gadge},
  \citenamefont {Pierz}, \citenamefont {Schumacher}, \citenamefont {Momeni},
  \citenamefont {Steil} \emph {et~al.}}]{merboldt2025observation}%
  \BibitemOpen
  \bibfield  {author} {\bibinfo {author} {\bibfnamefont {M.}~\bibnamefont
  {Merboldt}}, \bibinfo {author} {\bibfnamefont {M.}~\bibnamefont
  {Sch{\"u}ler}}, \bibinfo {author} {\bibfnamefont {D.}~\bibnamefont
  {Schmitt}}, \bibinfo {author} {\bibfnamefont {J.~P.}\ \bibnamefont {Bange}},
  \bibinfo {author} {\bibfnamefont {W.}~\bibnamefont {Bennecke}}, \bibinfo
  {author} {\bibfnamefont {K.}~\bibnamefont {Gadge}}, \bibinfo {author}
  {\bibfnamefont {K.}~\bibnamefont {Pierz}}, \bibinfo {author} {\bibfnamefont
  {H.~W.}\ \bibnamefont {Schumacher}}, \bibinfo {author} {\bibfnamefont
  {D.}~\bibnamefont {Momeni}}, \bibinfo {author} {\bibfnamefont
  {D.}~\bibnamefont {Steil}}, \emph {et~al.},\ }\href
  {https://doi.org/10.1038/s41567-025-02889-7} {\bibfield  {journal} {\bibinfo
  {journal} {Nat. Phys.}\ }\textbf {\bibinfo {volume} {21}},\ \bibinfo {pages}
  {1093} (\bibinfo {year} {2025})}\BibitemShut {NoStop}%
\bibitem [{\citenamefont {Oka}\ and\ \citenamefont
  {Aoki}(2009)}]{oka2009Photovoltaic}%
  \BibitemOpen
  \bibfield  {author} {\bibinfo {author} {\bibfnamefont {T.}~\bibnamefont
  {Oka}}\ and\ \bibinfo {author} {\bibfnamefont {H.}~\bibnamefont {Aoki}},\
  }\href {https://doi.org/10.1103/PhysRevB.79.081406} {\bibfield  {journal}
  {\bibinfo  {journal} {Phys. Rev. B}\ }\textbf {\bibinfo {volume} {79}},\
  \bibinfo {pages} {081406(R)} (\bibinfo {year} {2009})}\BibitemShut {NoStop}%
\bibitem [{\citenamefont {Fu}\ \emph {et~al.}(2026{\natexlab{a}})\citenamefont
  {Fu}, \citenamefont {Mondal}, \citenamefont {Liu}, \citenamefont {Tanaka},\
  and\ \citenamefont {Cayao}}]{Fu2026Floquet}%
  \BibitemOpen
  \bibfield  {author} {\bibinfo {author} {\bibfnamefont {P.-H.}\ \bibnamefont
  {Fu}}, \bibinfo {author} {\bibfnamefont {S.}~\bibnamefont {Mondal}}, \bibinfo
  {author} {\bibfnamefont {J.-F.}\ \bibnamefont {Liu}}, \bibinfo {author}
  {\bibfnamefont {Y.}~\bibnamefont {Tanaka}},\ and\ \bibinfo {author}
  {\bibfnamefont {J.}~\bibnamefont {Cayao}},\ }\href
  {https://doi.org/10.1103/lkf9-jgv6} {\bibfield  {journal} {\bibinfo
  {journal} {Phys. Rev. Lett.}\ }\textbf {\bibinfo {volume} {136}},\ \bibinfo
  {pages} {066703} (\bibinfo {year} {2026}{\natexlab{a}})}\BibitemShut
  {NoStop}%
\bibitem [{\citenamefont {Yarmohammadi}\ \emph
  {et~al.}(2026{\natexlab{b}})\citenamefont {Yarmohammadi}, \citenamefont
  {Fu},\ and\ \citenamefont {Freericks}}]{yarmohammadi2026efficient}%
  \BibitemOpen
  \bibfield  {author} {\bibinfo {author} {\bibfnamefont {M.}~\bibnamefont
  {Yarmohammadi}}, \bibinfo {author} {\bibfnamefont {P.-H.}\ \bibnamefont
  {Fu}},\ and\ \bibinfo {author} {\bibfnamefont {J.~K.}\ \bibnamefont
  {Freericks}},\ }\href@noop {} {\bibfield  {journal} {\bibinfo  {journal}
  {arXiv preprint arXiv:2602.20862}\ } (\bibinfo {year}
  {2026}{\natexlab{b}})}\BibitemShut {NoStop}%
\bibitem [{\citenamefont {Fu}\ \emph {et~al.}(2026{\natexlab{b}})\citenamefont
  {Fu}, \citenamefont {Mondal}, \citenamefont {Liu},\ and\ \citenamefont
  {Cayao}}]{Fu2026Light}%
  \BibitemOpen
  \bibfield  {author} {\bibinfo {author} {\bibfnamefont {P.-H.}\ \bibnamefont
  {Fu}}, \bibinfo {author} {\bibfnamefont {S.}~\bibnamefont {Mondal}}, \bibinfo
  {author} {\bibfnamefont {J.-F.}\ \bibnamefont {Liu}},\ and\ \bibinfo {author}
  {\bibfnamefont {J.}~\bibnamefont {Cayao}},\ }\href
  {https://doi.org/10.21468/SciPostPhys.20.2.059} {\bibfield  {journal}
  {\bibinfo  {journal} {SciPost Phys.}\ }\textbf {\bibinfo {volume} {20}},\
  \bibinfo {pages} {059} (\bibinfo {year} {2026}{\natexlab{b}})}\BibitemShut
  {NoStop}%
\bibitem [{\citenamefont {Yu}\ \emph {et~al.}(2021)\citenamefont {Yu},
  \citenamefont {Zhang},\ and\ \citenamefont {Song}}]{yu2021dynamical}%
  \BibitemOpen
  \bibfield  {author} {\bibinfo {author} {\bibfnamefont {J.}~\bibnamefont
  {Yu}}, \bibinfo {author} {\bibfnamefont {R.-X.}\ \bibnamefont {Zhang}},\ and\
  \bibinfo {author} {\bibfnamefont {Z.-D.}\ \bibnamefont {Song}},\ }\href
  {https://doi.org/10.1038/s41467-021-26092-3} {\bibfield  {journal} {\bibinfo
  {journal} {Nat. Commun.}\ }\textbf {\bibinfo {volume} {12}},\ \bibinfo
  {pages} {5985} (\bibinfo {year} {2021})}\BibitemShut {NoStop}%
\bibitem [{\citenamefont {Engelhardt}\ and\ \citenamefont
  {Cao}(2021)}]{Engelhardt2021Dynamical}%
  \BibitemOpen
  \bibfield  {author} {\bibinfo {author} {\bibfnamefont {G.}~\bibnamefont
  {Engelhardt}}\ and\ \bibinfo {author} {\bibfnamefont {J.}~\bibnamefont
  {Cao}},\ }\href {https://doi.org/10.1103/PhysRevLett.126.090601} {\bibfield
  {journal} {\bibinfo  {journal} {Phys. Rev. Lett.}\ }\textbf {\bibinfo
  {volume} {126}},\ \bibinfo {pages} {090601} (\bibinfo {year}
  {2021})}\BibitemShut {NoStop}%
\bibitem [{\citenamefont {Neufeld}\ \emph {et~al.}(2019)\citenamefont
  {Neufeld}, \citenamefont {Podolsky},\ and\ \citenamefont
  {Cohen}}]{neufeld2019floquet}%
  \BibitemOpen
  \bibfield  {author} {\bibinfo {author} {\bibfnamefont {O.}~\bibnamefont
  {Neufeld}}, \bibinfo {author} {\bibfnamefont {D.}~\bibnamefont {Podolsky}},\
  and\ \bibinfo {author} {\bibfnamefont {O.}~\bibnamefont {Cohen}},\ }\href
  {https://doi.org/10.1038/s41467-018-07935-y} {\bibfield  {journal} {\bibinfo
  {journal} {Nat. Commun.}\ }\textbf {\bibinfo {volume} {10}},\ \bibinfo
  {pages} {405} (\bibinfo {year} {2019})}\BibitemShut {NoStop}%
\bibitem [{\citenamefont {Xu}\ and\ \citenamefont {Wu}(2018)}]{Xu2018Space}%
  \BibitemOpen
  \bibfield  {author} {\bibinfo {author} {\bibfnamefont {S.}~\bibnamefont
  {Xu}}\ and\ \bibinfo {author} {\bibfnamefont {C.}~\bibnamefont {Wu}},\ }\href
  {https://doi.org/10.1103/PhysRevLett.120.096401} {\bibfield  {journal}
  {\bibinfo  {journal} {Phys. Rev. Lett.}\ }\textbf {\bibinfo {volume} {120}},\
  \bibinfo {pages} {096401} (\bibinfo {year} {2018})}\BibitemShut {NoStop}%
\bibitem [{\citenamefont {Ikeda}\ \emph {et~al.}(2023)\citenamefont {Ikeda},
  \citenamefont {Kitamura},\ and\ \citenamefont
  {Morimoto}}]{ikeda2023photocurrent}%
  \BibitemOpen
  \bibfield  {author} {\bibinfo {author} {\bibfnamefont {Y.}~\bibnamefont
  {Ikeda}}, \bibinfo {author} {\bibfnamefont {S.}~\bibnamefont {Kitamura}},\
  and\ \bibinfo {author} {\bibfnamefont {T.}~\bibnamefont {Morimoto}},\ }\href
  {https://doi.org/10.1103/PhysRevLett.131.096301} {\bibfield  {journal}
  {\bibinfo  {journal} {Phys. Rev. Lett.}\ }\textbf {\bibinfo {volume} {131}},\
  \bibinfo {pages} {096301} (\bibinfo {year} {2023})}\BibitemShut {NoStop}%
\bibitem [{\citenamefont {Trevisan}\ \emph {et~al.}(2022)\citenamefont
  {Trevisan}, \citenamefont {Arribi}, \citenamefont {Heinonen}, \citenamefont
  {Slager},\ and\ \citenamefont {Orth}}]{trevisan2022bicircular}%
  \BibitemOpen
  \bibfield  {author} {\bibinfo {author} {\bibfnamefont {T.~V.}\ \bibnamefont
  {Trevisan}}, \bibinfo {author} {\bibfnamefont {P.~V.}\ \bibnamefont
  {Arribi}}, \bibinfo {author} {\bibfnamefont {O.}~\bibnamefont {Heinonen}},
  \bibinfo {author} {\bibfnamefont {R.-J.}\ \bibnamefont {Slager}},\ and\
  \bibinfo {author} {\bibfnamefont {P.~P.}\ \bibnamefont {Orth}},\ }\href
  {https://doi.org/10.1103/PhysRevLett.128.066602} {\bibfield  {journal}
  {\bibinfo  {journal} {Phys. Rev. Lett.}\ }\textbf {\bibinfo {volume} {128}},\
  \bibinfo {pages} {066602} (\bibinfo {year} {2022})}\BibitemShut {NoStop}%
\bibitem [{\citenamefont {Mitra}\ \emph {et~al.}(2024)\citenamefont {Mitra},
  \citenamefont {Jim{\'e}nez-Gal{\'a}n}, \citenamefont {Aulich}, \citenamefont
  {Neuhaus}, \citenamefont {Silva}, \citenamefont {Pervak}, \citenamefont
  {Kling},\ and\ \citenamefont {Biswas}}]{mitra2024light}%
  \BibitemOpen
  \bibfield  {author} {\bibinfo {author} {\bibfnamefont {S.}~\bibnamefont
  {Mitra}}, \bibinfo {author} {\bibfnamefont {{\'A}.}~\bibnamefont
  {Jim{\'e}nez-Gal{\'a}n}}, \bibinfo {author} {\bibfnamefont {M.}~\bibnamefont
  {Aulich}}, \bibinfo {author} {\bibfnamefont {M.}~\bibnamefont {Neuhaus}},
  \bibinfo {author} {\bibfnamefont {R.~E.}\ \bibnamefont {Silva}}, \bibinfo
  {author} {\bibfnamefont {V.}~\bibnamefont {Pervak}}, \bibinfo {author}
  {\bibfnamefont {M.~F.}\ \bibnamefont {Kling}},\ and\ \bibinfo {author}
  {\bibfnamefont {S.}~\bibnamefont {Biswas}},\ }\href
  {https://doi.org/10.1038/s41586-024-07244-z} {\bibfield  {journal} {\bibinfo
  {journal} {Nature}\ }\textbf {\bibinfo {volume} {628}},\ \bibinfo {pages}
  {752} (\bibinfo {year} {2024})}\BibitemShut {NoStop}%
\bibitem [{\citenamefont {Tyulnev}\ \emph {et~al.}(2024)\citenamefont
  {Tyulnev}, \citenamefont {Jim{\'e}nez-Gal{\'a}n}, \citenamefont {Poborska},
  \citenamefont {Vamos}, \citenamefont {Russell}, \citenamefont {Tani},
  \citenamefont {Smirnova}, \citenamefont {Ivanov}, \citenamefont {Silva},\
  and\ \citenamefont {Biegert}}]{tyulnev2024valleytronics}%
  \BibitemOpen
  \bibfield  {author} {\bibinfo {author} {\bibfnamefont {I.}~\bibnamefont
  {Tyulnev}}, \bibinfo {author} {\bibfnamefont {{\'A}.}~\bibnamefont
  {Jim{\'e}nez-Gal{\'a}n}}, \bibinfo {author} {\bibfnamefont {J.}~\bibnamefont
  {Poborska}}, \bibinfo {author} {\bibfnamefont {L.}~\bibnamefont {Vamos}},
  \bibinfo {author} {\bibfnamefont {P.~S.~J.}\ \bibnamefont {Russell}},
  \bibinfo {author} {\bibfnamefont {F.}~\bibnamefont {Tani}}, \bibinfo {author}
  {\bibfnamefont {O.}~\bibnamefont {Smirnova}}, \bibinfo {author}
  {\bibfnamefont {M.}~\bibnamefont {Ivanov}}, \bibinfo {author} {\bibfnamefont
  {R.~E.}\ \bibnamefont {Silva}},\ and\ \bibinfo {author} {\bibfnamefont
  {J.}~\bibnamefont {Biegert}},\ }\href
  {https://doi.org/10.1038/s41586-024-07156-y} {\bibfield  {journal} {\bibinfo
  {journal} {Nature}\ }\textbf {\bibinfo {volume} {628}},\ \bibinfo {pages}
  {746} (\bibinfo {year} {2024})}\BibitemShut {NoStop}%
\bibitem [{\citenamefont {Lesko}\ \emph
  {et~al.}(2026{\natexlab{a}})\citenamefont {Lesko}, \citenamefont {Weitz},
  \citenamefont {Wittigschlager}, \citenamefont {Li}, \citenamefont {Heide},
  \citenamefont {Neufeld},\ and\ \citenamefont
  {Hommelhoff}}]{lesko2026optical}%
  \BibitemOpen
  \bibfield  {author} {\bibinfo {author} {\bibfnamefont {D.~M.}\ \bibnamefont
  {Lesko}}, \bibinfo {author} {\bibfnamefont {T.}~\bibnamefont {Weitz}},
  \bibinfo {author} {\bibfnamefont {S.}~\bibnamefont {Wittigschlager}},
  \bibinfo {author} {\bibfnamefont {W.}~\bibnamefont {Li}}, \bibinfo {author}
  {\bibfnamefont {C.}~\bibnamefont {Heide}}, \bibinfo {author} {\bibfnamefont
  {O.}~\bibnamefont {Neufeld}},\ and\ \bibinfo {author} {\bibfnamefont
  {P.}~\bibnamefont {Hommelhoff}},\ }\href
  {https://doi.org/0.1038/s41567-026-03429-7} {\bibfield  {journal} {\bibinfo
  {journal} {Nature Physics}\ ,\ \bibinfo {pages} {1}} (\bibinfo {year}
  {2026}{\natexlab{a}})}\BibitemShut {NoStop}%
\bibitem [{\citenamefont {Lesko}\ \emph
  {et~al.}(2026{\natexlab{b}})\citenamefont {Lesko}, \citenamefont {Weitz},
  \citenamefont {Wittigschlager}, \citenamefont {No{\"o}cker}, \citenamefont
  {Li}, \citenamefont {Hommelhoff},\ and\ \citenamefont
  {Neufeld}}]{lesko2026probing}%
  \BibitemOpen
  \bibfield  {author} {\bibinfo {author} {\bibfnamefont {D.~M.}\ \bibnamefont
  {Lesko}}, \bibinfo {author} {\bibfnamefont {T.}~\bibnamefont {Weitz}},
  \bibinfo {author} {\bibfnamefont {S.}~\bibnamefont {Wittigschlager}},
  \bibinfo {author} {\bibfnamefont {S.}~\bibnamefont {No{\"o}cker}}, \bibinfo
  {author} {\bibfnamefont {W.}~\bibnamefont {Li}}, \bibinfo {author}
  {\bibfnamefont {P.}~\bibnamefont {Hommelhoff}},\ and\ \bibinfo {author}
  {\bibfnamefont {O.}~\bibnamefont {Neufeld}},\ }\href@noop {} {\bibfield
  {journal} {\bibinfo  {journal} {ACS nano}\ }\textbf {\bibinfo {volume}
  {20}},\ \bibinfo {pages} {11614} (\bibinfo {year}
  {2026}{\natexlab{b}})}\BibitemShut {NoStop}%
\bibitem [{\citenamefont {Xiao}\ \emph {et~al.}(2024)\citenamefont {Xiao},
  \citenamefont {Zhao}, \citenamefont {Li}, \citenamefont {Shindou},\ and\
  \citenamefont {Song}}]{Xiao2024Spin}%
  \BibitemOpen
  \bibfield  {author} {\bibinfo {author} {\bibfnamefont {Z.}~\bibnamefont
  {Xiao}}, \bibinfo {author} {\bibfnamefont {J.}~\bibnamefont {Zhao}}, \bibinfo
  {author} {\bibfnamefont {Y.}~\bibnamefont {Li}}, \bibinfo {author}
  {\bibfnamefont {R.}~\bibnamefont {Shindou}},\ and\ \bibinfo {author}
  {\bibfnamefont {Z.-D.}\ \bibnamefont {Song}},\ }\href
  {https://doi.org/10.1103/PhysRevX.14.031037} {\bibfield  {journal} {\bibinfo
  {journal} {Phys. Rev. X}\ }\textbf {\bibinfo {volume} {14}},\ \bibinfo
  {pages} {031037} (\bibinfo {year} {2024})}\BibitemShut {NoStop}%
\bibitem [{\citenamefont {Jiang}\ \emph {et~al.}(2024)\citenamefont {Jiang},
  \citenamefont {Song}, \citenamefont {Zhu}, \citenamefont {Fang},
  \citenamefont {Weng}, \citenamefont {Liu}, \citenamefont {Yang},\ and\
  \citenamefont {Fang}}]{Jiang2024Enumeration}%
  \BibitemOpen
  \bibfield  {author} {\bibinfo {author} {\bibfnamefont {Y.}~\bibnamefont
  {Jiang}}, \bibinfo {author} {\bibfnamefont {Z.}~\bibnamefont {Song}},
  \bibinfo {author} {\bibfnamefont {T.}~\bibnamefont {Zhu}}, \bibinfo {author}
  {\bibfnamefont {Z.}~\bibnamefont {Fang}}, \bibinfo {author} {\bibfnamefont
  {H.}~\bibnamefont {Weng}}, \bibinfo {author} {\bibfnamefont {Z.-X.}\
  \bibnamefont {Liu}}, \bibinfo {author} {\bibfnamefont {J.}~\bibnamefont
  {Yang}},\ and\ \bibinfo {author} {\bibfnamefont {C.}~\bibnamefont {Fang}},\
  }\href {https://doi.org/10.1103/PhysRevX.14.031039} {\bibfield  {journal}
  {\bibinfo  {journal} {Phys. Rev. X}\ }\textbf {\bibinfo {volume} {14}},\
  \bibinfo {pages} {031039} (\bibinfo {year} {2024})}\BibitemShut {NoStop}%
\bibitem [{\citenamefont {Chen}\ \emph {et~al.}(2024)\citenamefont {Chen},
  \citenamefont {Ren}, \citenamefont {Zhu}, \citenamefont {Yu}, \citenamefont
  {Zhang}, \citenamefont {Liu}, \citenamefont {Li}, \citenamefont {Liu},
  \citenamefont {Li},\ and\ \citenamefont {Liu}}]{Chen2024Enumeration}%
  \BibitemOpen
  \bibfield  {author} {\bibinfo {author} {\bibfnamefont {X.}~\bibnamefont
  {Chen}}, \bibinfo {author} {\bibfnamefont {J.}~\bibnamefont {Ren}}, \bibinfo
  {author} {\bibfnamefont {Y.}~\bibnamefont {Zhu}}, \bibinfo {author}
  {\bibfnamefont {Y.}~\bibnamefont {Yu}}, \bibinfo {author} {\bibfnamefont
  {A.}~\bibnamefont {Zhang}}, \bibinfo {author} {\bibfnamefont
  {P.}~\bibnamefont {Liu}}, \bibinfo {author} {\bibfnamefont {J.}~\bibnamefont
  {Li}}, \bibinfo {author} {\bibfnamefont {Y.}~\bibnamefont {Liu}}, \bibinfo
  {author} {\bibfnamefont {C.}~\bibnamefont {Li}},\ and\ \bibinfo {author}
  {\bibfnamefont {Q.}~\bibnamefont {Liu}},\ }\href
  {https://doi.org/10.1103/PhysRevX.14.031038} {\bibfield  {journal} {\bibinfo
  {journal} {Phys. Rev. X}\ }\textbf {\bibinfo {volume} {14}},\ \bibinfo
  {pages} {031038} (\bibinfo {year} {2024})}\BibitemShut {NoStop}%
\bibitem [{\citenamefont {Jungwirth}\ \emph {et~al.}(2026)\citenamefont
  {Jungwirth}, \citenamefont {Sinova}, \citenamefont {Fernandes}, \citenamefont
  {Liu}, \citenamefont {Watanabe}, \citenamefont {Murakami}, \citenamefont
  {Nakatsuji},\ and\ \citenamefont {{\v{S}}mejkal}}]{jungwirth2026symmetry}%
  \BibitemOpen
  \bibfield  {author} {\bibinfo {author} {\bibfnamefont {T.}~\bibnamefont
  {Jungwirth}}, \bibinfo {author} {\bibfnamefont {J.}~\bibnamefont {Sinova}},
  \bibinfo {author} {\bibfnamefont {R.~M.}\ \bibnamefont {Fernandes}}, \bibinfo
  {author} {\bibfnamefont {Q.}~\bibnamefont {Liu}}, \bibinfo {author}
  {\bibfnamefont {H.}~\bibnamefont {Watanabe}}, \bibinfo {author}
  {\bibfnamefont {S.}~\bibnamefont {Murakami}}, \bibinfo {author}
  {\bibfnamefont {S.}~\bibnamefont {Nakatsuji}},\ and\ \bibinfo {author}
  {\bibfnamefont {L.}~\bibnamefont {{\v{S}}mejkal}},\ }\href
  {https://doi.org/10.1038/s41586-025-09883-2} {\bibfield  {journal} {\bibinfo
  {journal} {Nature}\ }\textbf {\bibinfo {volume} {649}},\ \bibinfo {pages}
  {837} (\bibinfo {year} {2026})}\BibitemShut {NoStop}%
\bibitem [{\citenamefont {Zeng}\ \emph {et~al.}(2026)\citenamefont {Zeng},
  \citenamefont {Qin}, \citenamefont {Qin}, \citenamefont {Feng}, \citenamefont
  {Wu}, \citenamefont {Xu},\ and\ \citenamefont {Wang}}]{zeng2026odd}%
  \BibitemOpen
  \bibfield  {author} {\bibinfo {author} {\bibfnamefont {M.}~\bibnamefont
  {Zeng}}, \bibinfo {author} {\bibfnamefont {Z.}~\bibnamefont {Qin}}, \bibinfo
  {author} {\bibfnamefont {L.}~\bibnamefont {Qin}}, \bibinfo {author}
  {\bibfnamefont {S.}~\bibnamefont {Feng}}, \bibinfo {author} {\bibfnamefont
  {L.}~\bibnamefont {Wu}}, \bibinfo {author} {\bibfnamefont {D.-H.}\
  \bibnamefont {Xu}},\ and\ \bibinfo {author} {\bibfnamefont {R.}~\bibnamefont
  {Wang}},\ }\href {https://doi.org/10.1103/7kmk-yl2t} {\bibfield  {journal}
  {\bibinfo  {journal} {Phys. Rev. B}\ }\textbf {\bibinfo {volume} {113}},\
  \bibinfo {pages} {L220412} (\bibinfo {year} {2026})}\BibitemShut {NoStop}%
\bibitem [{\citenamefont {Luo}\ \emph {et~al.}(2025)\citenamefont {Luo},
  \citenamefont {Hu}, \citenamefont {Hu},\ and\ \citenamefont
  {Law}}]{luo2025spin}%
  \BibitemOpen
  \bibfield  {author} {\bibinfo {author} {\bibfnamefont {X.-J.}\ \bibnamefont
  {Luo}}, \bibinfo {author} {\bibfnamefont {J.-X.}\ \bibnamefont {Hu}},
  \bibinfo {author} {\bibfnamefont {M.-L.}\ \bibnamefont {Hu}},\ and\ \bibinfo
  {author} {\bibfnamefont {K.}~\bibnamefont {Law}},\ }\href@noop {} {\bibfield
  {journal} {\bibinfo  {journal} {arXiv preprint arXiv:2510.05512}\ } (\bibinfo
  {year} {2025})}\BibitemShut {NoStop}%
\bibitem [{\citenamefont {Wang}\ \emph {et~al.}(2026)\citenamefont {Wang},
  \citenamefont {Yu}, \citenamefont {Cui}, \citenamefont {Han}, \citenamefont
  {He}, \citenamefont {Wu}, \citenamefont {Zhang}, \citenamefont {Yang},\ and\
  \citenamefont {Yao}}]{Wang2026Type}%
  \BibitemOpen
  \bibfield  {author} {\bibinfo {author} {\bibfnamefont {Y.}~\bibnamefont
  {Wang}}, \bibinfo {author} {\bibfnamefont {Z.-M.}\ \bibnamefont {Yu}},
  \bibinfo {author} {\bibfnamefont {C.}~\bibnamefont {Cui}}, \bibinfo {author}
  {\bibfnamefont {Y.}~\bibnamefont {Han}}, \bibinfo {author} {\bibfnamefont
  {T.}~\bibnamefont {He}}, \bibinfo {author} {\bibfnamefont {W.}~\bibnamefont
  {Wu}}, \bibinfo {author} {\bibfnamefont {R.-W.}\ \bibnamefont {Zhang}},
  \bibinfo {author} {\bibfnamefont {S.~A.}\ \bibnamefont {Yang}},\ and\
  \bibinfo {author} {\bibfnamefont {Y.}~\bibnamefont {Yao}},\ }\href
  {https://doi.org/10.1103/p3xh-f5bx} {\bibfield  {journal} {\bibinfo
  {journal} {Phys. Rev. Lett.}\ }\textbf {\bibinfo {volume} {136}},\ \bibinfo
  {pages} {106402} (\bibinfo {year} {2026})}\BibitemShut {NoStop}%
\bibitem [{\citenamefont {Mazin}\ \emph {et~al.}(2023)\citenamefont {Mazin},
  \citenamefont {Gonz{\'a}lez-Hern{\'a}ndez},\ and\ \citenamefont
  {{\v{S}}mejkal}}]{mazin2023induced}%
  \BibitemOpen
  \bibfield  {author} {\bibinfo {author} {\bibfnamefont {I.}~\bibnamefont
  {Mazin}}, \bibinfo {author} {\bibfnamefont {R.}~\bibnamefont
  {Gonz{\'a}lez-Hern{\'a}ndez}},\ and\ \bibinfo {author} {\bibfnamefont
  {L.}~\bibnamefont {{\v{S}}mejkal}},\ }\href@noop {} {\bibfield  {journal}
  {\bibinfo  {journal} {arXiv preprint arXiv:2309.02355}\ } (\bibinfo {year}
  {2023})}\BibitemShut {NoStop}%
\bibitem [{SM()}]{SM}%
  \BibitemOpen
  \href@noop {} {}\bibinfo {note} {See the Supplementary Material, which
  includes Refs. \cite{Kresse1996vasp, Kresse1999vasp,Perdew1996PBE,
  Blochl1994PBE,marzari1997maximally,
  souza2001maximally,Pizzi2020wannier90,WU2018WannierTools,zhi2022wannsymm,Park2014Interference},
  for the details of the first-principles calculations, the tight-binding
  models, the Floquet-Volkov sates for different light and the details of the
  spin point groups for collinear magnets.}\BibitemShut {Stop}%
\bibitem [{\citenamefont {Liu}\ \emph {et~al.}(2026{\natexlab{b}})\citenamefont
  {Liu}, \citenamefont {Yu}, \citenamefont {Zhang},\ and\ \citenamefont
  {Liu}}]{liu2026complete}%
  \BibitemOpen
  \bibfield  {author} {\bibinfo {author} {\bibfnamefont {Y.}~\bibnamefont
  {Liu}}, \bibinfo {author} {\bibfnamefont {J.}~\bibnamefont {Yu}}, \bibinfo
  {author} {\bibfnamefont {P.}~\bibnamefont {Zhang}},\ and\ \bibinfo {author}
  {\bibfnamefont {C.-C.}\ \bibnamefont {Liu}},\ }\href@noop {} {\bibfield
  {journal} {\bibinfo  {journal} {arXiv preprint arXiv:2607.19303}\ } (\bibinfo
  {year} {2026}{\natexlab{b}})}\BibitemShut {NoStop}%
\bibitem [{\citenamefont {Zhu}\ \emph {et~al.}(2026{\natexlab{c}})\citenamefont
  {Zhu}, \citenamefont {Yan},\ and\ \citenamefont
  {Yarmohammadi}}]{zhu2026parity}%
  \BibitemOpen
  \bibfield  {author} {\bibinfo {author} {\bibfnamefont {D.}~\bibnamefont
  {Zhu}}, \bibinfo {author} {\bibfnamefont {Z.}~\bibnamefont {Yan}},\ and\
  \bibinfo {author} {\bibfnamefont {M.}~\bibnamefont {Yarmohammadi}},\
  }\href@noop {} {\bibfield  {journal} {\bibinfo  {journal} {arXiv preprint
  arXiv:2607.27335}\ } (\bibinfo {year} {2026}{\natexlab{c}})}\BibitemShut
  {NoStop}%
\bibitem [{\citenamefont {Kresse}\ and\ \citenamefont
  {Furthm\"uller}(1996)}]{Kresse1996vasp}%
  \BibitemOpen
  \bibfield  {author} {\bibinfo {author} {\bibfnamefont {G.}~\bibnamefont
  {Kresse}}\ and\ \bibinfo {author} {\bibfnamefont {J.}~\bibnamefont
  {Furthm\"uller}},\ }\href {https://doi.org/10.1103/PhysRevB.54.11169}
  {\bibfield  {journal} {\bibinfo  {journal} {Phys. Rev. B}\ }\textbf {\bibinfo
  {volume} {54}},\ \bibinfo {pages} {11169} (\bibinfo {year}
  {1996})}\BibitemShut {NoStop}%
\bibitem [{\citenamefont {Kresse}\ and\ \citenamefont
  {Joubert}(1999)}]{Kresse1999vasp}%
  \BibitemOpen
  \bibfield  {author} {\bibinfo {author} {\bibfnamefont {G.}~\bibnamefont
  {Kresse}}\ and\ \bibinfo {author} {\bibfnamefont {D.}~\bibnamefont
  {Joubert}},\ }\href {https://doi.org/10.1103/PhysRevB.59.1758} {\bibfield
  {journal} {\bibinfo  {journal} {Phys. Rev. B}\ }\textbf {\bibinfo {volume}
  {59}},\ \bibinfo {pages} {1758} (\bibinfo {year} {1999})}\BibitemShut
  {NoStop}%
\bibitem [{\citenamefont {Perdew}\ \emph {et~al.}(1996)\citenamefont {Perdew},
  \citenamefont {Burke},\ and\ \citenamefont {Ernzerhof}}]{Perdew1996PBE}%
  \BibitemOpen
  \bibfield  {author} {\bibinfo {author} {\bibfnamefont {J.~P.}\ \bibnamefont
  {Perdew}}, \bibinfo {author} {\bibfnamefont {K.}~\bibnamefont {Burke}},\ and\
  \bibinfo {author} {\bibfnamefont {M.}~\bibnamefont {Ernzerhof}},\ }\href
  {https://doi.org/10.1103/PhysRevLett.77.3865} {\bibfield  {journal} {\bibinfo
   {journal} {Phys. Rev. Lett.}\ }\textbf {\bibinfo {volume} {77}},\ \bibinfo
  {pages} {3865} (\bibinfo {year} {1996})}\BibitemShut {NoStop}%
\bibitem [{\citenamefont {Bl\"ochl}(1994)}]{Blochl1994PBE}%
  \BibitemOpen
  \bibfield  {author} {\bibinfo {author} {\bibfnamefont {P.~E.}\ \bibnamefont
  {Bl\"ochl}},\ }\href {https://doi.org/10.1103/PhysRevB.50.17953} {\bibfield
  {journal} {\bibinfo  {journal} {Phys. Rev. B}\ }\textbf {\bibinfo {volume}
  {50}},\ \bibinfo {pages} {17953} (\bibinfo {year} {1994})}\BibitemShut
  {NoStop}%
\bibitem [{\citenamefont {Marzari}\ and\ \citenamefont
  {Vanderbilt}(1997)}]{marzari1997maximally}%
  \BibitemOpen
  \bibfield  {author} {\bibinfo {author} {\bibfnamefont {N.}~\bibnamefont
  {Marzari}}\ and\ \bibinfo {author} {\bibfnamefont {D.}~\bibnamefont
  {Vanderbilt}},\ }\href {https://doi.org/10.1103/PhysRevB.56.12847} {\bibfield
   {journal} {\bibinfo  {journal} {Phys. Rev. B}\ }\textbf {\bibinfo {volume}
  {56}},\ \bibinfo {pages} {12847} (\bibinfo {year} {1997})}\BibitemShut
  {NoStop}%
\bibitem [{\citenamefont {Souza}\ \emph {et~al.}(2001)\citenamefont {Souza},
  \citenamefont {Marzari},\ and\ \citenamefont
  {Vanderbilt}}]{souza2001maximally}%
  \BibitemOpen
  \bibfield  {author} {\bibinfo {author} {\bibfnamefont {I.}~\bibnamefont
  {Souza}}, \bibinfo {author} {\bibfnamefont {N.}~\bibnamefont {Marzari}},\
  and\ \bibinfo {author} {\bibfnamefont {D.}~\bibnamefont {Vanderbilt}},\
  }\href {https://doi.org/10.1103/PhysRevB.65.035109} {\bibfield  {journal}
  {\bibinfo  {journal} {Phys. Rev. B}\ }\textbf {\bibinfo {volume} {65}},\
  \bibinfo {pages} {035109} (\bibinfo {year} {2001})}\BibitemShut {NoStop}%
\bibitem [{\citenamefont {Pizzi}\ \emph {et~al.}(2020)\citenamefont {Pizzi},
  \citenamefont {Vitale}, \citenamefont {Arita}, \citenamefont {Bl{\"u}gel},
  \citenamefont {Freimuth}, \citenamefont {G{\'e}ranton}, \citenamefont
  {Gibertini}, \citenamefont {Gresch}, \citenamefont {Johnson}, \citenamefont
  {Koretsune} \emph {et~al.}}]{Pizzi2020wannier90}%
  \BibitemOpen
  \bibfield  {author} {\bibinfo {author} {\bibfnamefont {G.}~\bibnamefont
  {Pizzi}}, \bibinfo {author} {\bibfnamefont {V.}~\bibnamefont {Vitale}},
  \bibinfo {author} {\bibfnamefont {R.}~\bibnamefont {Arita}}, \bibinfo
  {author} {\bibfnamefont {S.}~\bibnamefont {Bl{\"u}gel}}, \bibinfo {author}
  {\bibfnamefont {F.}~\bibnamefont {Freimuth}}, \bibinfo {author}
  {\bibfnamefont {G.}~\bibnamefont {G{\'e}ranton}}, \bibinfo {author}
  {\bibfnamefont {M.}~\bibnamefont {Gibertini}}, \bibinfo {author}
  {\bibfnamefont {D.}~\bibnamefont {Gresch}}, \bibinfo {author} {\bibfnamefont
  {C.}~\bibnamefont {Johnson}}, \bibinfo {author} {\bibfnamefont
  {T.}~\bibnamefont {Koretsune}}, \emph {et~al.},\ }\href
  {https://doi.org/10.1088/1361-648x/ab51ff} {\bibfield  {journal} {\bibinfo
  {journal} {J. Phys. Condens. Matter}\ }\textbf {\bibinfo {volume} {32}},\
  \bibinfo {pages} {165902} (\bibinfo {year} {2020})}\BibitemShut {NoStop}%
\bibitem [{\citenamefont {Wu}\ \emph {et~al.}(2018)\citenamefont {Wu},
  \citenamefont {Zhang}, \citenamefont {Song}, \citenamefont {Troyer},\ and\
  \citenamefont {Soluyanov}}]{WU2018WannierTools}%
  \BibitemOpen
  \bibfield  {author} {\bibinfo {author} {\bibfnamefont {Q.}~\bibnamefont
  {Wu}}, \bibinfo {author} {\bibfnamefont {S.}~\bibnamefont {Zhang}}, \bibinfo
  {author} {\bibfnamefont {H.-F.}\ \bibnamefont {Song}}, \bibinfo {author}
  {\bibfnamefont {M.}~\bibnamefont {Troyer}},\ and\ \bibinfo {author}
  {\bibfnamefont {A.~A.}\ \bibnamefont {Soluyanov}},\ }\href
  {https://doi.org/10.1016/j.cpc.2017.09.033} {\bibfield  {journal} {\bibinfo
  {journal} {Comput. Phys. Commun.}\ }\textbf {\bibinfo {volume} {224}},\
  \bibinfo {pages} {405 } (\bibinfo {year} {2018})}\BibitemShut {NoStop}%
\bibitem [{\citenamefont {Zhi}\ \emph {et~al.}(2022)\citenamefont {Zhi},
  \citenamefont {Xu}, \citenamefont {Wu}, \citenamefont {Ning},\ and\
  \citenamefont {Cao}}]{zhi2022wannsymm}%
  \BibitemOpen
  \bibfield  {author} {\bibinfo {author} {\bibfnamefont {G.-X.}\ \bibnamefont
  {Zhi}}, \bibinfo {author} {\bibfnamefont {C.}~\bibnamefont {Xu}}, \bibinfo
  {author} {\bibfnamefont {S.-Q.}\ \bibnamefont {Wu}}, \bibinfo {author}
  {\bibfnamefont {F.}~\bibnamefont {Ning}},\ and\ \bibinfo {author}
  {\bibfnamefont {C.}~\bibnamefont {Cao}},\ }\href
  {https://doi.org/10.1016/j.cpc.2021.108196} {\bibfield  {journal} {\bibinfo
  {journal} {Comput. Phys. Commun.}\ }\textbf {\bibinfo {volume} {271}},\
  \bibinfo {pages} {108196} (\bibinfo {year} {2022})}\BibitemShut {NoStop}%
\bibitem [{\citenamefont {Park}(2014)}]{Park2014Interference}%
  \BibitemOpen
  \bibfield  {author} {\bibinfo {author} {\bibfnamefont {S.~T.}\ \bibnamefont
  {Park}},\ }\href {https://doi.org/10.1103/PhysRevA.90.013420} {\bibfield
  {journal} {\bibinfo  {journal} {Phys. Rev. A}\ }\textbf {\bibinfo {volume}
  {90}},\ \bibinfo {pages} {013420} (\bibinfo {year} {2014})}\BibitemShut
  {NoStop}%
\end{thebibliography}%


\onecolumngrid

\vspace{0.5em}
\begin{center}
    \textbf{ END MATTER}
\end{center}
\vspace{0.5em}

\twocolumngrid

\setcounter{equation}{0}
\renewcommand{\theequation}{A\arabic{equation}}

\begin{table*}[t]
\centering
\caption{{Light-driven spin splitting in two-dimensional systems}. The 2D spin layer groups are organized according to their undriven spin-splitting type and the resulting classes under CPL, BLPL with $\alpha=0$, and BLPL with $\alpha=\pi/4$. AFM denotes a spin-degenerate antiferromagnetic class, whereas FiM denotes a ferrimagnetic-like class in which no remaining symmetry relates the two spin sectors. We align the $a$-axis of  lattice along the $x$-direction. There are 63 collinear spin layer point groups for compensated collinear magnets, and the numbering of the spin layer point group  follows Ref.~\cite{Wang2026Type}.}
\label{tab:2d}
\scriptsize
\setlength{\tabcolsep}{3pt}
\renewcommand{\arraystretch}{1.08}
\begin{tabular*}{\textwidth}{@{\extracolsep{\fill}}lllll@{}}
\toprule
\textbf{No light} & \textbf{CPL} & \textbf{BLPL, $\alpha=0$} & \textbf{BLPL, $\alpha=\pi/4$} & \textbf{Groups} \\
\midrule
\multirow{5}{*}{AFM} & AFM & AFM & AFM & 4, 6, 10, 13, 16, 20, 22, 23, 27, 28, 36, 37, 40, 47, 51, 53, 57, 58, 59, 62, 63 \\
 & p-wave & d-wave & mixed-parity & 7, 14, 21 \\
 & p-wave & FiM & FiM & 1, 2, 8, 9, 18 \\
 & f-wave & d-wave & mixed-parity & 46, 50, 60 \\
 & f-wave & FiM & FiM & 41, 45, 48, 52, 55 \\
\midrule
\multirow{4}{*}{d-wave} & d-wave & d-wave & mixed-parity & 11, 12, 17, 19, 32, 34, 38 \\
 & d-wave & FiM & FiM & 24, 25, 26, 30, 35 \\
 & mixed-parity & d-wave & mixed-parity & 5, 15 \\
 & mixed-parity & FiM & FiM & 3 \\
\midrule
\multirow{1}{*}{g-wave} & g-wave & d-wave & mixed-parity & 29, 31, 33, 39 \\
\midrule
\multirow{3}{*}{i-wave} & i-wave & d-wave & mixed-parity & 44, 49, 54, 61 \\
 & mixed-parity & d-wave & mixed-parity & 43, 56 \\
 & mixed-parity & FiM & FiM & 42 \\
\bottomrule
\end{tabular*}
\end{table*}

\begin{table*}[t]
\centering
\caption{{Light-driven spin-splitting classes in three-dimensional systems. The 3D spin point groups are organized according to their undriven spin-splitting type and the resulting classes under CPL, BLPL with $\alpha=0$, and BLPL with $\alpha=\pi/4$.} AFM denotes a spin-degenerate antiferromagnetic class, whereas FiM denotes a ferrimagnetic-like class in which no remaining symmetry relates the two spin sectors. We align the $a$-axis of  lattice along the $x$-direction. 
In contrast to 2D systems, the 3D classification also includes higher-order $h$- and $k$-wave spin splittings. The numbering of the spin point group  follows Ref.~\cite{Wang2026Type}.}
\label{tab:3d}
\scriptsize
\setlength{\tabcolsep}{3pt}
\renewcommand{\arraystretch}{1.08}
\begin{tabular*}{\textwidth}{@{\extracolsep{\fill}}lllll@{}}
\toprule
\textbf{No light} & \textbf{CPL} & \textbf{BLPL, $\alpha=0$} & \textbf{BLPL, $\alpha=\pi/4$} & \textbf{Groups} \\
\midrule
AFM &
\begin{tabular}{@{}l@{}}
p-wave\\
p-wave\\
f-wave\\
f-wave\\
h-wave\\
k-wave
\end{tabular}
&
\begin{tabular}{@{}l@{}}
d-wave\\
FiM\\
d-wave\\
FiM\\
d-wave\\
d-wave
\end{tabular}
&
\begin{tabular}{@{}l@{}}
mixed-parity\\
FiM\\
mixed-parity\\
FiM\\
mixed-parity\\
mixed-parity
\end{tabular}
&
\begin{tabular}{@{}l@{}}
4, 11, 12, 17, 29, 35, 42, 52, 53, 58\\
1, 5, 30\\
16, 25, 49, 57\\
34, 41\\
26\\
51
\end{tabular}
\\ \midrule
d-wave &
\begin{tabular}{@{}l@{}}
d-wave\\
d-wave\\
mixed-parity\\
mixed-parity
\end{tabular}
&
\begin{tabular}{@{}l@{}}
d-wave\\
FiM\\
d-wave\\
FiM
\end{tabular}
&
\begin{tabular}{@{}l@{}}
mixed-parity\\
FiM\\
mixed-parity\\
FiM
\end{tabular}
&
\begin{tabular}{@{}l@{}}
6,10,27\\
15\\
3,7,8,9,21,23\\
2,13,14,19,24
\end{tabular}
\\ \midrule
g-wave &
\begin{tabular}{@{}l@{}}
g-wave\\
mixed-parity\\
mixed-parity
\end{tabular}
&
\begin{tabular}{@{}l@{}}
d-wave\\
d-wave\\
FiM
\end{tabular}
&
\begin{tabular}{@{}l@{}}
mixed-parity\\
mixed-parity\\
FiM
\end{tabular}
&
\begin{tabular}{@{}l@{}}
28, 33, 40, 48\\
18, 20, 22, 32, 36, 39, 46, 47\\
31, 37, 44
\end{tabular}
\\ \midrule
i-wave &
\begin{tabular}{@{}l@{}}
d-wave\\
i-wave\\
mixed-parity\\
mixed-parity
\end{tabular}
&
\begin{tabular}{@{}l@{}}
FiM\\
d-wave\\
d-wave\\
FiM
\end{tabular}
&
\begin{tabular}{@{}l@{}}
FiM\\
mixed-parity\\
mixed-parity\\
FiM
\end{tabular}
&
\begin{tabular}{@{}l@{}}
56\\
50\\
38, 43, 45\\
54, 55
\end{tabular}
\\
\bottomrule
\end{tabular*}
\end{table*}

\emph{Appendix A: Dynamical symmetry of the light-driven Hamiltonian}-
To verify Eq. (\ref{Eq-invariant}), we apply the dynamical spin-group operation  $\gamma=[S||R]\cdot \tau_\delta $, represented as  $\hat{D}_\gamma = \hat{D}_{[S||R]} \cdot {\tau}_\delta $, to the time-dependent Hamiltonian:
\begin{equation}
\begin{split}
&\hat{D}_\gamma H_s\left(\mathbf{k}, t\right) \hat{D}_\gamma^{-1} \\
& =\hat{D}_{{[S||R]}} H_s\left[\mathbf{k}+e\mathbf{A}\left(\eta_S t-\eta_{S}\delta\right)/\hbar\right] \hat{D}_{{[S||R]}}^{-1} \\
& =H_{\chi_S s}\left[\eta_S R \mathbf{k}+e \eta_S R \mathbf{A}\left(\eta_S t-\eta_{S}\delta\right)/\hbar\right] \\
& =H_{\chi_S s}\left[\eta_S R \mathbf{k}+e\mathbf{A}(t)/\hbar\right] \\
& =H_{\chi_S s}\left(\eta_S R \mathbf{k}, t\right).
\end{split}
\label{transformation_Dg}
\end{equation}
Here, $\eta_S= \pm 1$ is defined in the main text, with $\eta_S=-1$ for an antiunitary spin-space operation, which is associated with time reversal. In going from the third to the fourth line, we have used the dynamical symmetry of the vector potential of light [Eq.~(\ref{eq:light_symm})]. If we denote $g=[S||R]$, $s_{g}=\chi_{S}s$, $\mathbf{k}_g=\eta_SR\mathbf{k}$, and $\eta_{g}=\eta_S$, Eq.~(\ref{transformation_Dg}) can be notionally  simplified as 
\begin{equation}
\begin{split}
H_{s_g}\left(\mathbf{k}_g, t\right)=\hat{D}_g H_s\left(\mathbf{k}, \eta_g t-\eta_g\delta\right) \hat{D}_g^{-1}.
\end{split}
\label{transformation_Dg2}
\end{equation}

\textit{Appendix B: Dynamical-symmetry constrains on quasienergy spectrum and Floquet wavefunctions}--Expanding Eq. (\ref{EQ-floquetEQ}) 
in Fourier harmonics, the eigenvalue problem becomes  
\begin{equation}
\begin{split}
\mathcal{H}^F_s(\mathbf{k})|u_{s,\beta }(\mathbf{k})\rangle\rangle=\varepsilon_{s,\beta } (\mathbf{k})|u_{s,\beta }(\mathbf{k})\rangle\rangle,
\end{split}
\end{equation}
where 
\begin{equation}
\begin{split}
\left[\mathcal{H}^F_s(\mathbf{k})\right]_{p q}=H_s^{(p-q)}(\mathbf{k})+p \hbar \omega \delta_{p q},
\end{split}
\label{samble_ham}
\end{equation}
with $H_s^{(m)}(\mathbf{k})=\frac{1}{T} \int_0^T d t H_s(\mathbf{k}, t) e^{-\mathrm{i} m \omega t}$ and the eigenstates $\left.\left|u_{s\beta\mathbf{k}}\right\rangle\right\rangle  $  are written in the extended Hilbert space (Sambe space) as
\begin{equation}
\begin{split}
\left.\left|u_{s,\beta}(\mathbf{k})\right\rangle\right\rangle=\left(\begin{array}{ccccc}
\dots,\left|u_{s,\beta}^{(-1)}(\mathbf{k})\right\rangle, \left|u_{s,\beta }^{(0)}(\mathbf{k})\right\rangle,\left|u_{s,\beta  }^{(1)}(\mathbf{k})\right\rangle,\dots\\
\end{array}\right)^T,
\end{split}
\end{equation}
with $\left|u_{s,\beta}^{(m)}(\mathbf{k})\right\rangle =\frac{1}{T} \int_0^T d t e^{-\mathrm{i} m \omega t} \left|u_{s,\beta}(\mathbf{k},t)\right\rangle$ .


By using Eq.~(\ref{transformation_Dg2}), it is easy to see the Fourier components of the Hamiltonian satisfy 
\begin{equation}
\begin{split}
& H^{(m)}_{s_g}(\mathbf{k}_g) \\
&=\frac{1}{T} \int_0^T d t H_{s_g}\left(\mathbf{k}_g, t\right)e^{-\mathrm{i} m \omega t} \\
&= \frac{1}{T} \int_0^T d t  \hat{D}_g H_s\left(\mathbf{k}, \eta_g(t-\delta)\right) \hat{D}_g^{-1} e^{-\mathrm{i} m \omega( t-\delta )} e^{-\mathrm{i} m \omega \delta}\\
&= \frac{1}{T} \int_0^T d t  \hat{D}_g H_s\left(\mathbf{k}, \eta_g(t-\delta)\right) e^{-\mathrm{i} m \omega \eta_g(t-\delta)} \hat{D}_g^{-1} e^{-\mathrm{i} m \omega \delta} \\
&=\hat{D}_g H_s^{(m)}(\mathbf{k}) \hat{D}_g^{-1}e^{-\mathrm{i} m \omega \delta} .
\end{split}
\label{transformation_Fourier}
\end{equation}
Here, in going from the third to the fourth line, we have considered if $\hat{D}_g^{-1}$ is antiunitary when moving the factor $e^{-\mathrm{i} m \omega( t-\delta )}$ across $\hat{D}_g^{-1}$. 

Next, we define the symmetry representation in Sambe space as
\begin{equation}
\begin{split}
\hat{\mathcal{G}}(g) & =\sum_{m \in \mathbb{Z}} e^{-\mathrm{i} m \omega \delta}|m\rangle\langle m| \otimes \hat{D}_g,
\end{split}
\end{equation}
where $m$ labels the Floquet harmonic and $|m\rangle\langle m|$ projects onto the corresponding Fourier sector.  Applying $\hat{\mathcal{G}}(g)$ to the Floquet Hamiltonian and using Eq.~(\ref{transformation_Fourier}), we get 
\begin{equation}
\begin{split}
\mathcal{H}_{s_g}^F\left(\mathbf{k}_g\right)=\mathcal{G}_{\mathrm{g}} \mathcal{H}_s^F(\mathbf{k}) \mathcal{G}_{\mathrm{g}}^{-1}.
\end{split}
\end{equation}
Therefore, for the nondegenrate band, the quasienergy satisfies 
 \begin{equation}
 \begin{split}
 \varepsilon_{s_g,\beta }(\mathbf{k}_g)=\varepsilon_{s,\beta}(\mathbf{k}) \quad(\bmod  \hbar \omega).
 \end{split}
 \end{equation}
which is independent of the time translation part of the dynamical symmetry operation. Consequently, when discussing the spin splitting of the quasienergy bands, it is sufficient to focus on the spin point group part. The classifications of light-induced spin splitting in 2D and 3D are summarized in the Table~\ref{tab:2d} and Table~\ref{tab:3d}, repsectively.

On the other hand,  for a nondegenerate band, the eigenvector satisfies 
\begin{equation}
\begin{split}
|u_{{s}_g,\beta}(\mathbf{k}_g)\rangle\rangle=e^{\mathrm{i} \gamma_{\beta}(\mathbf{k})} \mathcal{G}_{\mathrm{g}}|u_{s,\beta}(\mathbf{k})\rangle\rangle,
\end{split}
\end{equation}
where $\gamma_{\beta}$ is a gauge phase, which has no effect on the observables and can therefore be omitted. The harmonic components of the eigenvector satisfy
\begin{equation}
\begin{split}
\left|u_{s_g,\beta}^{(m)}(\mathbf{k}_g)\right\rangle= e^{-\mathrm{i}  m \omega \delta} \hat{D}_g\left|u_{s,\beta}^{(m)}(\mathbf{k})\right\rangle.
\end{split}
\end{equation}
Finally, the Floquet Schrödinger wavefunction is 
\begin{equation}
\begin{split}
\left|\Psi_{s,\beta}(\mathbf{k},t)\right\rangle&=e^{-\mathrm{i} \varepsilon_{s,\beta }(\mathbf{k}) t / \hbar}\left|u_{s,\beta}(\mathbf{k},t)\right\rangle\\
&=\sum_m e^{-\mathrm{i}\left[\varepsilon_{s,\beta}(\mathbf{k})-m \hbar \omega\right] t / \hbar}\left|u_{s,\beta}^{(m)}(\mathbf{k})\right\rangle,
\end{split}
\end{equation}
and it satisfies
\begin{equation}
\begin{split}
\left|\Psi_{s_g,\beta }(\mathbf{k}_g,t)\right\rangle= \sum_{m \in \mathbb{Z}} e^{-i m \omega \delta}  \times e^{-i\left[\varepsilon_{s,\beta }(\mathbf{k})-m \hbar \omega\right] t / \hbar} \hat{D}_g\left|u_{s,\beta }^{(m)}(\mathbf{k})\right\rangle.
\end{split}
\label{wavefunction_time}
\end{equation}

\end{document}